\documentclass[10pt,aps,pra,twocolumn,floatfix,nofootinbib,superscriptaddress,longbibliography]{revtex4-2}

\usepackage{bm,graphicx,mathrsfs,amsmath,amssymb,mathtools,makecell,bbm, amsthm,dsfont,color,times,txfonts,nicefrac,framed,enumitem,tikz,physics,wrapfig,amsfonts,lipsum,xargs,xcolor,comment}
\definecolor{boxcolor}{RGB}{215,215,253}

\usepackage[breakable]{tcolorbox}

\newtcolorbox{mybox}[1]{breakable,%
  breakable,%
  colback=boxcolor!5!white,%
  colframe=boxcolor!75!black,%
  fonttitle=\bfseries,%
  title=#1%
}

\definecolor{changescolor}{RGB}{180,80,30}

\usepackage[dvipsnames]{xcolor}
\usepackage[colorinlistoftodos,prependcaption,textsize=tiny]{todonotes}
\newcommandx{\unsure}[2][1=]{\todo[linecolor=red,backgroundcolor=red!25,bordercolor=red,#1]{#2}}
\newcommandx{\change}[2][1=]{\todo[linecolor=blue,backgroundcolor=blue!25,bordercolor=blue,#1]{#2}}
\newcommandx{\info}[2][1=]{\todo[linecolor=OliveGreen,backgroundcolor=OliveGreen!25,bordercolor=OliveGreen,#1]{#2}}
\newcommandx{\improvement}[2][1=]{\todo[linecolor=OliveGreen,backgroundcolor=Plum!25,bordercolor=Plum,#1]{#2}}
\newcommandx{\thiswillnotshow}[2][1=]{\todo[disable,#1]{#2}}

\usepackage[colorlinks=true,linkcolor=alecolor,citecolor=alecolor]{hyperref}

\hypersetup{urlcolor=alecolor}
\definecolor{equationcolor}{RGB}{222,94,100}
\definecolor{alecolor}{RGB}{198,113,190}
\definecolor{changescolor}{rgb}{0, 0, 0.7}

\newenvironment{sproof}{%
  \proof}{\endproof}

\makeatletter
\def\blfootnote{\gdef\@thefnmark{}\@footnotetext}
\makeatother

\renewcommand{\v}[1]{\ensuremath{\boldsymbol #1}}
\newcommand{\ms}[1]{\textsf{#1}}

\newcommand{\iden}{\mathbbm{1}}

\newcommand{\E}[1]{\mathcal{E}}

\def\E{ {\cal E} }

\newtheorem{thm}{Theorem}

\newtheorem{lem}[thm]{Lemma}
\newtheorem{prop}[thm]{Proposition}
\newtheorem{cor}[thm]{Corollary}

\newtheorem{defn}{Definition}
\newcommand{\sean}[1]{{\color{purple}}}

\newtheorem*{thmrestated}{Theorem~\ref{thm:no-rank-one-average-gaussian-main} restated}

\begin{document}

\title{Privacy in continuous-variable distributed quantum sensing}

\author{A. de Oliveira Junior}
\email{alexssandredeoliveira@gmail.com}
    \affiliation{Center for Macroscopic Quantum States bigQ, Department of Physics,
Technical University of Denmark, Fysikvej 307, 2800 Kgs. Lyngby, Denmark}
\author{Anton L. Andersen}
 \affiliation{Sorbonne Université, CNRS, LIP6, F-75005 Paris, France}
\author{Benjamin Lundgren Larsen}
    \affiliation{Center for Macroscopic Quantum States bigQ, Department of Physics,
Technical University of Denmark, Fysikvej 307, 2800 Kgs. Lyngby, Denmark}
\author{Sean William Moore}
\affiliation{Sorbonne Université, CNRS, LIP6, F-75005 Paris, France}
\author{Damian Markham}
\affiliation{Sorbonne Université, CNRS, LIP6, F-75005 Paris, France}
\author{Masahiro Takeoka}
    \affiliation{Department of Electronics and Electrical Engineering, Keio University, 3-14-1 Hiyoshi, Kohoku-ku, Yokohama 223-8522, Japan}
    \affiliation{Advanced ICT Research Institute, National Institute of Information and Communications Technology (NICT), Koganei, Tokyo 184-8795, Japan}
\author{Jonatan Bohr Brask}
\author{Ulrik L. Andersen}
	\affiliation{Center for Macroscopic Quantum States bigQ, Department of Physics,
Technical University of Denmark, Fysikvej 307, 2800 Kgs. Lyngby, Denmark}
\date{\today}

\begin{abstract}
Can a distributed network of quantum sensors estimate a global parameter while protecting every locally encoded value? We answer this question affirmatively by introducing and analysing a protocol for distributed quantum sensing in the continuous-variable regime. We consider a multipartite network in which an unknown local phase is imprinted at each node on a shared entangled Gaussian state. We show that the average phase can be estimated with high precision, exhibiting Heisenberg scaling in the total photon number, while individual phases are inaccessible. We further prove a no-go theorem showing that, for three or more parties, no finite-energy Gaussian probe can provide complete privacy of the average phase, meaning that all phase combinations orthogonal to the average remain entirely hidden. This identifies the two-party case as an exceptional Gaussian setting that can achieve complete privacy. We further investigate the impact of displacements and optical losses, revealing trade-offs between estimation accuracy and privacy. Finally, we benchmark the protocol against other continuous-variable resource states.
\end{abstract}

\maketitle

\section{Introduction \label{Sec:introduction}}

Sensing spatially distributed parameters lies at the heart of emerging quantum technologies, with implications ranging from clock synchronisation~\cite{komar2014quantum,Dai2020} to distributed computing~\cite{Barz2012,Fitzsimons2017} and precision metrology~\cite{Giovannetti2006,Zhuang2018}. Over the past two decades, we have learned how quantum information may be processed, shared, and protected across networks~\cite{Giovannetti2002,Kimble2008,Gottesman2012,Baumgratz2016,Liu2020}. Within this landscape, quantum sensor networks, in which entangled resources are distributed among spatially separated nodes that collaboratively estimate parameters, have emerged as powerful platforms for both fundamental studies and real-world applications~\cite{proctor2017networkedquantumsensing,Proctor2018,Ge2018,Zhuang2018,Rubio2020,Sekatski2020,Wlk2020,Zhang2021,Hamann2024}.

A central challenge in any such network is privacy, meaning that each node should learn only the authorised global quantity while remaining ignorant of the individual parameters. A paradigmatic example is distributed phase sensing, where each node encodes an unknown local phase and the network aims to estimate only the average phase, while keeping all individual phases and non-target phase combinations hidden~\cite{shettell2022, Hassani2025}. In this setting, each local phase is an unknown physical parameter imprinted at a node by a local channel. The party operating that node follows the prescribed measurement procedure, but does not know the value of its phase except through the quantum probe and the public measurement record. Likewise, a party or external observer may access the full public classical record and any quantum systems assigned to them, but has no direct side information about the raw local phase values.  Operationally, this setting captures applications ranging from networked atomic clocks, where the encoded phase represents a local clock offset and the target is an average synchronisation signal~\cite{komar2014quantum,Dai2020}, to distributed magnetic-field or strain sensing, where local fields are encoded as phases but only an aggregate quantity is revealed~\cite{Sekatski2020,Hamann2024,Proctor2018}. It also encompasses more general quantum sensor networks estimating linear functions of spatially distributed parameters~\cite{Ge2018,Rubio2020,Zhuang2018,guo2020distributed}, as well as quantum-assisted long-baseline interferometry, where separated sites jointly estimate optical phase information without each site accessing the full global signal~\cite{Gottesman2012,khabiboulline2019optical,khabiboulline2019quantum}.

While discrete-variable quantum networks have protocols for private parameter estimation~\cite{shettell2022,Unnikrishnan2022,Moore2025}, including experimental demonstrations~\cite{ho2024quantumprivatedistributedsensing}, optimal state constructions~\cite{Hassani2025}, and analyses of robustness against noise and loss~\cite{Bugalho2025,ueda2025quantumnetworksensingefficient}, the corresponding question for continuous-variable (CV) sensor networks remains largely unexplored. In particular, it is not known whether CV systems can achieve complete privacy, whether weaker forms of privacy are possible, or how these privacy requirements affect the precision with which the target function can be estimated. This gap is especially striking because optical Gaussian states provide deterministic, large-scale entanglement at room temperature~\cite{Adesso2007,Krauter2013}. This motivates the key question addressed in the present work: under an experimentally feasible protocol, can CV sensor networks be made private and, if so, what price must be paid in precision?

\begin{figure}[t]
    \centering
    \includegraphics{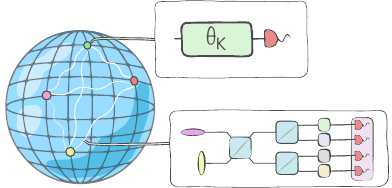}
    \caption{\emph{Phase estimation \& privacy}. Two squeezed states are combined at a local beam splitter to generate a two-mode squeezed state, which is then split into four modes using two additional beam splitters. Each party receives one mode, and a local phase $\theta_k$ is encoded. The goal is to estimate the average phase while hiding the individual local phases. The strongest version would hide every combination orthogonal to the average (complete privacy). For Gaussian probes of three or more modes complete privacy is ruled out, so the multipartite protocol targets the privacy of each local phase (weak privacy) and quantifies residual leakage into other combinations.}
    \label{F-scheme}
\end{figure}

We answer this question by proving a no-go theorem showing that, for three or more parties, finite-energy Gaussian states cannot hide every phase combination orthogonal to the average phase, and therefore cannot provide complete privacy. The two-party case is the exceptional setting allowed by our result, where a two-mode squeezed state can achieve complete privacy. For larger networks, the attainable goal is instead privacy at the level of individual local phases. We then identify a simple, experimentally feasible Gaussian protocol, based on distributing a two-mode squeezed vacuum through a beam-splitter network, that preserves the privacy of every individual local phase and asymptotically approaches complete privacy (see Fig.~\ref{F-scheme} for a schematic representation). We also show that the average phase can be estimated with high precision, exhibiting Heisenberg scaling in the total photon number, while other orthogonal phase combinations exhibit shot-noise scaling. Our results reveal a fundamental limitation on the use of Gaussian resources for privacy in distributed quantum sensing. They also complement recent work~\cite{alushi2026privacydistributedquantumsensing} by resolving the open question of whether Gaussian states can achieve complete privacy.

\begin{figure*}
    \centering
    \includegraphics{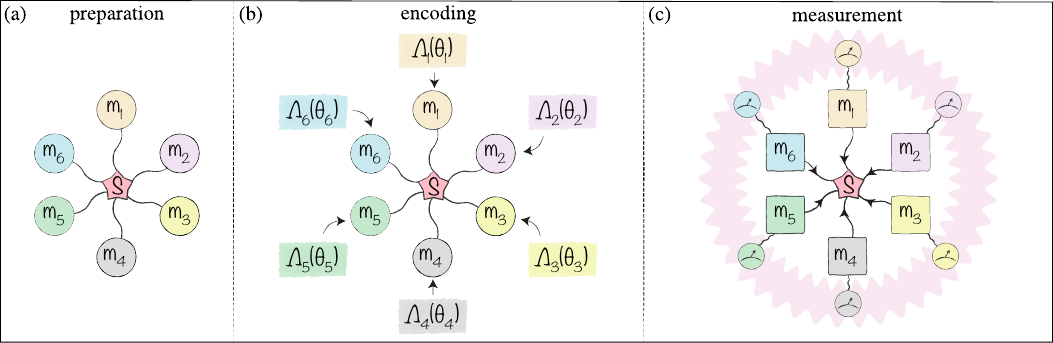}
    \caption{\emph{Estimation scheme}. The parties (depicted as coloured circles) request an $M$-partite state $\rho_0$ from the source $\ms{S}$ (red star), which (a) prepares and distributes it within the network. Then, (b) an unknown parameter $\theta_{\mu}$ is locally encoded via a quantum operation $\Lambda_{\mu}(\theta_{\mu})$. After encoding, the probe is measured either locally at the nodes, with the outcomes broadcast publicly, or by a joint receiver, and the resulting data are processed by a trusted central processor.}
    \label{F-estimation-scheme}
\end{figure*}

The paper is organised as follows. In Sec.~\ref{Sec:setting-the-scene}, we introduce the estimation task, the notion of privacy, the QFIm-based privacy criterion, and the Gaussian formalism used throughout. There, we also derive a general QFIm formula for pure Gaussian states under local phase encoding in Lemma~\ref{Lem:QFIm}, together with a covariance-matrix expression for the privacy measure in Theorem~\ref{Thm:privacy-n}. In Sec.~\ref{Sec:privacy-in-CV}, we first prove the no-go theorem for complete average-phase privacy with three or more Gaussian modes in Theorem~\ref{thm:no-rank-one-average-gaussian-main}. We then present the two-mode squeezed state as the exceptional two-party case achieving complete privacy, including the effects of finite squeezing and loss, before introducing the multipartite beam-splitter protocol. For this protocol, we characterise the QFIm in Lemma~\ref{lem:tree-QFIM-characterisation}, derive its spectrum in Theorem~\ref{thm:tree-QFIM-spectrum}, and analyse its privacy properties in Corollary~\ref{cor:privacy-weak-not-strong}. In Sec.~\ref{Sec:privacy-optimality}, we study the effects of displacement and optical loss, and benchmark the protocol against other Gaussian resources, including independent single-mode squeezed states and cluster states. We conclude with a summary and outlook, while the appendices contain the technical proofs and supporting examples.

\section{Setting the scene}\label{Sec:setting-the-scene}

This section sets out the estimation problem, the estimation-theoretic tools we use, the operational notions of privacy, and the Gaussian framework in which we work. We first introduce the multiparameter setting and the quantum Fisher information matrix, whose range and kernel determine which linear combinations of phases are locally estimable. We then define unobservable directions, private parameters, and the privacy regimes considered in the rest of the paper, together with the assumptions on the parties under which these guarantees apply. Finally, we specialise the formalism to Gaussian states and operations.

\subsection{Multiparameter estimation problem}\label{Sec:multiparameter}

We consider a network of $M$ parties, connected by quantum and classical links, that share an initial joint quantum state $\hat\rho$. Each party $\mu$ is subjected to a local encoding channel $\Lambda_\mu(\theta_\mu)$ that imprints an unknown parameter $\theta_\mu\in\mathbb R$ on its share (see Fig.~\hyperref[F-estimation-scheme]{\ref{F-estimation-scheme}a}). The resulting encoded state is $\hat\rho(\v\theta) := \v\Lambda_{\v\theta}(\hat\rho)$, where $\v \Lambda_{\v\theta} := \bigotimes_{\mu=1}^{M}\Lambda_\mu(\theta_\mu)$, with $\v\theta=(\theta_1,\dots,\theta_M)\in\mathbb R^M$ (see Fig.~\hyperref[F-estimation-scheme]{\ref{F-estimation-scheme}b}). After encoding, every node performs a measurement and broadcasts its classical outcome to a trusted central processor (see Fig.~\hyperref[F-estimation-scheme]{\ref{F-estimation-scheme}c}). The collective task is to estimate, from these broadcast outcomes, a publicly agreed linear target function $f(\v\theta) = \v u^{\ms T}\v\theta$ where $\v u\in\mathbb R^M$ and $\|\v u\|_2 = 1$.

The fundamental object that bounds the precision of any unbiased estimator of $\v\theta$ is the quantum Fisher information matrix (QFIm), with entries~\cite{Paris2009,Dominik2018,Liu2019,Adresso2021}
\begin{equation}\label{Eq:QFIm}
    \v Q_{\mu\nu}[\hat\rho(\v\theta)] := \frac{1}{2}\tr \bigl[\hat\rho(\v\theta)\,\{\hat L_{\theta_\mu},\hat L_{\theta_\nu}\}\bigr],
\end{equation}
where $\{\hat A,\hat B\} = \hat A\hat B + \hat B\hat A$ denotes the anticommutator and $2\partial_{\theta_\nu}\hat\rho(\v\theta) = \hat\rho(\v\theta)\hat L_{\theta_\nu} + \hat L_{\theta_\nu}\hat\rho(\v\theta)$ is the symmetric logarithmic derivative (SLD). The QFIm $\v Q$ is a real, symmetric, positive semi-definite matrix. Its spectrum governs the achievable estimation precision and, as we will see, also the information that can leak about $\v\theta$. We will use two subspaces of $\mathbbm{R}^M$:
\begin{equation}
\operatorname{range}(\v Q) := \{\v Q\v u : \v u\in\mathbb R^M\}\subseteq\mathbb R^M,
\end{equation}
i.e., the linear span of the columns of $\v Q$. The \emph{kernel} (or null space) is
\begin{equation}
\ker\v Q := \{\v u\in\mathbb R^M : \v Q\v u = \v 0\},
\end{equation}
equivalently, the subspace of vectors that $\v Q$ maps to zero. For symmetric $\v Q$, $\operatorname{range}(\v Q)$ and $\ker\v Q$ are orthogonal complements in $\mathbb R^M$. 

A linear function $\v u^{\ms T}\v\theta$ admits a finite-variance, locally unbiased estimator if and only if $\v u\in\operatorname{range}(\v Q)$. When this holds, the scalar quantum Cram\'er--Rao bound is~\cite{Ragy2016,Szczykulska03072016}
\begin{equation}\label{Eq:scalar-QCRB}
\operatorname{var}(\v u^{\ms T}\hat{\v\theta}) \;\ge\; \v u^{\ms T}\v Q^{+}\v u, 
\end{equation}
where $\v Q^{+}$ is the Moore--Penrose pseudoinverse. The condition $\v u\in\operatorname{range}(\v Q)$ is essential. If $\v u$ has a component along $\ker\v Q$, then that component lies in an unobservable direction. To make this explicit, decompose $\v u$ orthogonally as $\v u=\v u_r+\v u_0$, with $\v u_r\in\operatorname{range}(\v Q)$ and $\v u_0\in\ker\v Q$. For $\epsilon>0$, we regularise the QFIm as $\v Q+\epsilon\mathbbm 1$, and denote $\v Q_r$ the restriction of $\v Q$ to $\operatorname{range}(\v Q)$. Since $\operatorname{range}(\v Q)$ and $\ker\v Q$ are orthogonal invariant subspaces of $\v Q$, one obtains
\begin{equation}\label{Eq:scalar-QCRB-reg} 
\v u^{\ms T}(\v Q+\epsilon\mathbbm 1)^{-1}\v u = \v u_r^{\ms T}(\v Q_r+\epsilon\mathbbm 1)^{-1}\v u_r + \frac{\|\v u_0\|_2^2}{\epsilon}, 
\end{equation} 
which diverges as $\epsilon\to 0^{+}$ whenever $\v u_0\neq 0$. Any component of $\v u$ in $\ker\v Q$ is therefore \emph{unobservable}. No locally unbiased estimator of the corresponding linear combination of $\v\theta$ can have finite variance. For the unitary encodings considered in this work an even stronger statement holds. As shown below, $\v u\in\ker\v Q$ implies the exact invariance $\hat\rho(\v\theta+t\v u)=\hat\rho(\v\theta)$ for all $t\in\mathbb R$, so every measurement produces statistics that are strictly independent of the parameter combination along $\v u$, and that combination is nonidentifiable rather than merely poorly estimable (see Definition~\ref{Def:privacy-definition} and Lemma~\ref{Lem:equivalence} in Appendix~\ref{App:privacy-def}). We use $\v Q^{+}$ in subsequent expressions only as shorthand on $\operatorname{range}(\v Q)$, and we always identify a kernel direction explicitly as unobservable rather than as having vanishing variance. Thus, when $\v u\notin\operatorname{range}(\v Q)$, the finite number $\v u^{\ms T}\v Q^+\v u$ is not interpreted as a precision bound for the full target $\v u^{\ms T}\v\theta$. These observations motivate the following definition.
\begin{defn}[Unobservable direction \& private parameter]\label{Def:privacy-definition}
Let $\v Q(\v\theta)$ be the QFIm of the encoded state $\hat\rho(\v\theta)$, and define the directional derivative 
\begin{equation}
    \partial_{\v u}\hat\rho(\v\theta) := \sum_{\mu=1}^{M} u_\mu\, \partial_{\theta_\mu}\hat\rho(\v\theta).
\end{equation}
\begin{enumerate}
\item A nonzero vector $\v u\in\mathbb R^M$ is \emph{unobservable} at $\v\theta$ if $\v u\in\ker\v Q(\v\theta)$.
\item A component $\theta_j$ is \emph{private} at $\v\theta$ if there exists an unobservable $\v u$ with $u_j\neq 0$.
\end{enumerate}
\end{defn}
Equivalently, $\theta_j$ is private if and only if $\v e_j\notin\operatorname{range}(\v Q)$, where $\v e_j$ is the unit vector with a $1$ in the $j$-th entry (see Proposition~\ref{Prop:privacy-lemma} below). We note that the definition of unobservability, $\v u\in\ker\v Q$, coincides with $\partial_{\v u}\hat\rho=0$ at regular points (see Lemma~\ref{Lem:equivalence} in Appendix~\ref{App:privacy-def}), where $\hat\rho(\v\theta)$ has locally constant rank and depends smoothly on $\v\theta$~\cite{seveso2019,Liu2019}. Outside such points, parameter-dependent changes in the rank of $\hat\rho$ can produce singular Fisher-information contributions through the limiting behaviour of $\tfrac{(\partial_{\v u}p_k)^2}{p_k}$, where $p_k(\v\theta)$ is an eigenvalue of $\hat\rho$ satisfying $p_k(\v\theta_0)=\partial_{\v u}p_k(\v\theta_0)=0$ at an interior point $\v\theta_0$~\cite{Gefen2019,gardner2025}. Although both the numerator and denominator vanish at $\v\theta_0$, the limiting ratio along $\v u$ can be finite and nonzero. In the protocols studied here the parameter dependence is purely unitary, so the spectrum of $\hat\rho(\v\theta)$ is independent of $\v\theta$ and $\tfrac{\partial p_k}{\partial\theta_\mu}\equiv0$ for every eigenvalue (including zero eigenvalues). The rank of $\hat\rho(\v\theta)$ is locally constant, and no singular eigenvalue contribution arises. The only non-unitary channel we model---a fixed loss before the phase encoding---is itself $\v\theta$-independent and so does not introduce parameter-dependent support either. The regular-point equivalence between $\v u\in\ker\v Q$ and $\partial_{\v u}\hat\rho=0$ therefore holds throughout this work.

We conclude this section with a proposition that characterises private parameters:
\begin{prop}[Characterisation of private parameters]\label{Prop:privacy-lemma}
For a real symmetric QFIm $\v Q$, a component $\theta_j$ is private at $\v\theta$ if and only if $\v e_j\notin\operatorname{range}(\v Q)$. Equivalently, $\theta_j$ is private if and only if there exists $\v u\in\ker\v Q$ with $u_j\neq0$.
\end{prop}
\begin{proof}
Because $\v Q$ is real and symmetric, $\operatorname{range}(\v Q)=(\ker\v Q)^\perp$. If there exists $\v u\in\ker\v Q$ with $u_j\neq0$, then $\v e_j^{\ms T}\v u=u_j\neq0$, so $\v e_j$ is not orthogonal to $\ker\v Q$ and therefore cannot belong to $\operatorname{range}(\v Q)$. Conversely, if $\v e_j\notin\operatorname{range}(\v Q)$, then its orthogonal projection $\v u$ onto $\ker\v Q$ is nonzero. Since $\v e_j-\v u\in\operatorname{range}(\v Q)$ and $\v u\in\ker\v Q$, these two vectors are orthogonal. Hence $0=\v u^{\ms T}(\v e_j-\v u)=u_j-\|\v u\|_2^2$, so $u_j=\|\v u\|_2^2>0$. Thus $\v u\in\ker\v Q$ and $u_j\neq0$. By Definition~\ref{Def:privacy-definition}, this is exactly the condition that $\theta_j$ is private.
\end{proof}

\subsection{Privacy characterisation and quantification}\label{Sec:privacy-quantification}

We can now state the operational notions of privacy that organise the rest of the paper, together with the assumptions on the parties under which our results hold. Each party holds one mode of an entangled probe distributed by a trusted source. The parameter $\theta_\mu$ is an unknown physical phase imprinted on mode $\mu$ by the local environment--a sample, an external field, a refractive-index shift, a clock offset, or another local channel. Each party operates the post-encoding readout apparatus at its node and follows the prescribed measurement procedure, but does not know the numerical value of $\theta_\mu$ except through the quantum probe and the public measurement record. The privacy task is to ensure that no party, colluding subset of parties, or external observer with access to the public broadcast and to the quantum systems in their possession can uniquely reconstruct any raw individual value, except insofar as it is unavoidably constrained by the authorised target function.

We work under the following prescribed-protocol assumptions:
\begin{enumerate}[label=$\textbf{A}_{\arabic*.}$]
    \item \emph{Trusted source.} A trusted source prepares the requested probe within a known tolerance and distributes it through the network. The only state imperfection we model is optical loss in the distribution channels.
    \item \emph{Access-controlled \& no-side-information.} The parameter $\theta_\mu$ is an unknown external physical quantity imprinted once on the designated mode during a prescribed sensing interaction at node $\mu$. Party $\mu$ operates the post-encoding readout, but is neither given $\theta_\mu$ nor able to substitute an auxiliary probe or obtain an additional use of the parameter-dependent interaction.
    \item \emph{Prescribed-protocol execution.} The parties perform the agreed measurement and report its outcome truthfully.
    \item \emph{Available information.} A party, a colluding subset of nodes, or an external observer may have access to \emph{(i)} the entire public classical record, namely all broadcast measurement outcomes, and \emph{(ii)} the quantum systems in their possession. They do not have direct access to any raw local parameter value $\theta_\nu$, including the value encoded at their own node, except through the quantum systems in their possession and the public classical record.
\end{enumerate}
The privacy guarantees below are information-theoretic statements about the impossibility of uniquely inferring the protected parameter values from the accessible information. The restriction in $\mathbf A_2$ is a physical access assumption, not a property enforced by the shared probe. If a party could substitute an auxiliary probe or repeat the parameter-dependent interaction, no shared state could prevent it from estimating $\theta_\mu$ directly. Assumption $\mathbf A_3$ concerns only the stated estimation performance and the integrity of the reported estimate. For privacy, the parties may process arbitrarily the quantum systems in their possession and the public record, as specified in $\mathbf A_4$. The privacy guarantee does not cover additional access to the sensing interaction contrary to $\mathbf A_2$ or direct side information about the raw parameters, and is not a composable cryptographic guarantee. The only integrity-relevant imperfection we model is optical loss in the distribution channels. Relaxing these conditions using verification protocols for untrusted sources~\cite{Unnikrishnan2022,descamps2024} is discussed in the conclusions.

Here and throughout, $\v Q$ denotes the QFIm of the full encoded state, without imposing an LOCC restriction on the measurement. Conclusions concerning complete and component privacy that are derived from $\ker\v Q$ are measurement independent: for the unitary phase encodings considered here, a kernel direction leaves the encoded state invariant and is therefore unobservable under any POVM. By contrast, Eq.~\eqref{Eq:scalar-QCRB} and the nonzero eigenvalues of $\v Q$ provide unrestricted-measurement precision benchmarks. Their attainability by local measurements is not assumed for general $M$.

We now organise privacy in terms of the range and kernel of the QFIm. Fix a target direction $\v u\in\mathbb R^M$ with $\|\v u\|_2=1$. We exclude the operationally degenerate single-component case $\v u=\pm\v e_j$, since asking the protocol to reveal $\theta_j$ while simultaneously keeping it private is inconsistent. Complete privacy is the strongest case, in which only the target function $f(\v\theta)=\v u^{\ms T}\v\theta$ is accessible. A weaker requirement is component privacy, where every individual phase remains private even though some non-target linear combinations may still be accessible. For non-coordinate targets, complete privacy implies component privacy, but the converse need not hold.

To obtain a quantitative measure, rather than only a binary distinction between private and non-private directions, we use the measure of Ref.~\cite{Bugalho2025}:
\begin{equation}\label{Eq:label-privacy-measure}
P(\v Q,\v u) = \frac{\v u^{\ms T}\v Q\v u}{\tr(\v Q)}.
\end{equation}
The privacy measure is evaluated with $\|\v u\|_2=1$, so that $0\le P(\v Q,\v u)\le 1$ whenever $\v Q\neq\v 0$. It is undefined when $\tr\v Q=0$, which corresponds to the encoded state being locally invariant under all single-parameter shifts. Moreover, $P(\v Q,\v u)=1$ if and only if $\v Q$ is rank one and $\operatorname{range}(\v Q)=\operatorname{span}\{\v u\}$, equivalently $\v Q\propto \v u\v u^{\ms T}$. The measure has two convenient properties: it is basis invariant, $P(\v B\v Q\v B^{\ms T},\v B\v u)=P(\v Q,\v u)$ for any real orthogonal matrix $\v B$, and it is continuous in $\v Q$, so small experimental imperfections lead to proportionally small changes in $P$. Finally, $P(\v Q,\v u)>0$ does not by itself imply that the target $f(\v\theta)$ is estimable. This can happen because $\v u$ may have a component in $\ker\v Q$, which does not contribute to $\v u^{\ms T}\v Q\v u$ but nevertheless prevents $\v u^{\ms T}\hat{\v\theta}$ from having finite variance.

The scalar quantity $P(\v Q,\v u)$ is useful for quantifying how concentrated the Fisher information is along the target direction, but it does not by itself determine which individual parameters are private. We therefore keep track of the kernel and range of $\v Q$ explicitly. In the following classification, we assume $\v Q\succeq0$, $\tr\v Q>0$, a non-degenerate target $\v u\neq\pm\v e_j$, and local estimability of the target, $\v u\in\operatorname{range}(\v Q)$. Cases with $\v u\notin\operatorname{range}(\v Q)$ are excluded, since then the target itself is unobservable. Under these assumptions, the regimes below distinguish complete privacy, weak component privacy, partial hiding, and the absence of component privacy.
\begin{enumerate}
    \item \emph{Complete privacy} ($\operatorname{range}(\v Q)=\operatorname{span}\{\v u\}$, equivalently $P=1$ and $\dim\ker\v Q=M-1$): only $f(\v\theta)$ is accessible.
    \item \emph{Weak component privacy} ($\v e_j\notin\operatorname{range}(\v Q)$ for every $j$, but $\operatorname{range}(\v Q)\supsetneq\operatorname{span}\{\v u\}$): every individual component is private, but some non-target combinations remain accessible.
    \item \emph{Partial hiding} ($\dim\ker\v Q\ge 1$, but $\v e_j\in\operatorname{range}(\v Q)$ for at least one $j$): some linear combinations are unobservable, but at least one individual component $\theta_j$ is accessible.
    \item \emph{No component privacy} ($\ker\v Q=\{\v 0\}$): every component $\theta_j$ is observable.
\end{enumerate}

The following box illustrates these notions on a few discrete-variable warm-ups; analogous Gaussian examples are worked out in Sec.~\ref{Sec:privacy-in-CV}.

\begin{mybox}{{Example: privacy measure for GHZ, W \& $\ket{+}^{\otimes 3}$} }

As a warm-up, consider three parties sharing a given state $\ket{\psi}$. Each party $j$ locally encodes a phase $\theta_j$ by applying the unitary operation $U(\theta_j) = \ketbra{0}{0} + e^{i \theta_j} \ketbra{1}{1}$. The aim is to estimate the average phase, characterised by a vector $\v v_\text{avg} \propto (1,1,1)$, while remaining completely agnostic about other combinations of the individual phases. The privacy measure for different choices of the initial state (see Appendix~\ref{App-simple-example} for details) is given by
\begin{enumerate}[label=(\alph*), leftmargin=*, itemsep=2pt]
\item \!\!$\ket{\mathrm{GHZ}} \!=\! \frac{1}{\sqrt{2}}(\ket{000}+\ket{111}) \longrightarrow P(\v Q_{\textrm{GHZ}},\v v_\text{avg}) = 1.$
\end{enumerate}
Here $\v Q_{\text{GHZ}}$ is the all-ones matrix. Its kernel is the two-dimensional subspace $\{\v u:\ u_1+u_2+u_3=0\}$, so for each $j$ there exist null vectors with $u_j\neq0$. Hence every individual phase $\theta_j$ is private. The only observable combination is the sum $\theta_1+\theta_2+\theta_3$, which coincides with the chosen target. 
\begin{enumerate}[label=(\alph*), leftmargin=*, itemsep=2pt,start=2]
\item \!\!$\ket{\mathrm{W}} \!=\! \frac{1}{\sqrt{3}}(\ket{100}+\ket{010}+\ket{001})\!\longrightarrow\! P(\v Q_{W},\v v_\text{avg}) = 0.$
\end{enumerate}
One has $\ker\v Q_{\mathrm{W}}=\mathrm{span}\{(1,1,1)\}$ (any global shift of all three phases is invisible), so $\v Q_{\mathrm{W}}$ has rank two. Because the null vector has all three nonzero components, each $\theta_j$ is private. However, the average-phase direction lies entirely in the kernel, so $\v v_{\rm avg}\notin\operatorname{range}(\v Q_{\mathrm{W}})$ and the average phase is unobservable. The W state is therefore target-blind for the average. The privacy measure registers this as $P(\v Q_{\mathrm W},\v v_\text{avg})=0$.
\begin{enumerate}[label=(\alph*), leftmargin=*, itemsep=2pt,start=3]
\item $\ket{\text{+}}^{\otimes 3} = \qty[\frac{1}{\sqrt{2}}(\ket{0}+\ket{1})]^{\otimes 3} \longrightarrow P(\v Q_{+},\v v_\text{avg}) = \frac{1}{3}.$
\end{enumerate}
In this case, $\v Q_+$ is full rank (no kernel), so there is no null direction with $u_j\neq0$. By Definition~\ref{Def:privacy-definition} no individual phase is private. The probe reveals all three phases (and thus the average), with $P(\v Q_{+},\v v_\text{avg})=\tfrac13$ indicating that only one third of the total information is aligned with the average and the rest leaks into orthogonal combinations.  
\end{mybox}

\subsection{Gaussian framework}\label{Sec:Gaussian-framework}

We now restrict the framework above to Gaussian states and operations, which form the resource class used in the rest of the paper.

We consider quantum systems with $M$ canonical degrees of freedom, with annihilation and creation operators $\hat a_\mu$ and $\hat a^\dagger_\mu$ ($\mu\in\{1,\dots,M\}$). Setting $\hbar=1$, the canonical commutation relations are $[\hat a_\mu,\hat a_\nu]=[\hat a^\dagger_\mu,\hat a^\dagger_\nu]=0$ and $[\hat a_\mu,\hat a^\dagger_\nu]=\delta_{\mu\nu}$. Hermitian quadratures are defined by $\hat x_\mu=\tfrac{1}{\sqrt 2}(\hat a_\mu+\hat a^\dagger_\mu)$ and $\hat p_\mu=\tfrac{1}{i\sqrt 2}(\hat a_\mu-\hat a^\dagger_\mu)$, satisfying $[\hat x_\mu,\hat p_\nu]=i\delta_{\mu\nu}$. Collecting them in the column vector $\hat{\v r}=(\hat x_1,\hat p_1,\dots,\hat x_M,\hat p_M)^{\ms T}$, one has $[\hat{\v r},\hat{\v r}^{\ms T}]=i\Omega$, where $\Omega$ is the symplectic form. Throughout, $M\in\mathbb N$ denotes the number of bosonic modes (and the number of users); the auxiliary symbol $N\in\mathbb N$ appears only in recursive constructions, e.g.\ $M=2^N$.

A Gaussian state is completely characterised by its first moments and covariance matrix~\cite{Weedbrook2012,brask2022gaussianstatesoperations,serafini2023quantum},
\begin{equation}
\bar{\v r} = \tr(\hat\rho\,\hat{\v r}) \quad \text{and} \quad 
\v\sigma = \tfrac{1}{2}\tr\bigl[\hat\rho\{(\hat{\v r}-\bar{\v r}),(\hat{\v r}-\bar{\v r})^{\ms T}\}\bigr].
\end{equation}
Gaussian operations act on $(\v\sigma,\bar{\v r})$ via $\v\sigma\to\v F\v\sigma\v F^\top$, $\bar{\v r}\to\v F\bar{\v r}+\v d$ with $\v F\Omega\v F^\top=\Omega$. For pure Gaussian states, the QFIm can be written as~\cite{monras2013phasespaceformalismquantum}
\begin{equation}\label{Eq:QFI-cv}
\v Q_{\mu\nu}(\v\sigma,\bar{\v r}) = \tfrac{1}{4} \tr\left[ (\v\sigma^{-1}\partial_{\theta_\mu}\v\sigma) (\v\sigma^{-1}\partial_{\theta_\nu}\v\sigma) \right] + \partial_{\theta_\mu}\bar{\v r}^{\ms T} \v\sigma^{-1} \partial_{\theta_\nu}\bar{\v r}.
\end{equation}
The inverse $\v\sigma^{-1}$ here refers to the covariance matrix only. Every physical Gaussian state satisfies $\v\sigma\succ 0$, so $\v\sigma^{-1}$ is always well defined. This should not be confused with full rank of the density operator, which fails for pure Gaussian states. The first term in Eq.~\eqref{Eq:QFI-cv} encodes the information stored in the covariances, while the second accounts for the displacements; either term vanishes when the corresponding quantity is parameter-independent.

In the Gaussian protocols studied below, the local parameters are encoded as phase shifts generated by the number operators $\hat n_\mu=\hat a_\mu^\dagger\hat a_\mu$. The encoded family is
\begin{equation}\label{Eq:local-phase-encoding}
\hat\rho(\v\theta)=\hat U(\v\theta)\hat\rho\hat U(\v\theta)^\dagger
\:\:\text{where \:\:}
\hat U(\v\theta)=\bigotimes_{\mu=1}^{M} e^{i\theta_\mu \hat n_\mu}.
\end{equation}
For a generic pure multimode Gaussian state subjected to the local phase encoding in Eq.~\eqref{Eq:local-phase-encoding}, the quantum Fisher information matrix can be expressed directly in terms of the covariance matrix and first moments. The following result gives the corresponding closed-form expression:

\begin{lem}[QFIm for Gaussian phase shifts]\label{Lem:QFIm} Let $|\Psi(\v{\theta})\rangle$ be a pure $M$-mode Gaussian state obtained by applying local phase shifts to a pure state $\ket{\psi_0}$
\begin{equation}
   \ket{\Psi(\v{\theta})} = \left( \bigotimes_{\mu=1}^M e^{i \theta_\mu \hat{n}_\mu} \right) |\psi_0\rangle,
\end{equation}
The quantum Fisher information matrix associated with multi-parameter phase estimation $\v{\theta} = (\theta_1, \dots, \theta_M)$ can be expressed in terms of the quadrature covariance matrix and first moments as follows:
\begin{equation}
   \v Q_{\mu\nu} = 2 \sum_{z,w \in \{x, p\}} \left( \sigma_{\hat{z}_\mu \hat{w}_\nu}^2 + 2\, \sigma_{\hat{z}_\mu \hat{w}_\nu} \, \bar{\v{r}}_{\hat{z}_\mu} \bar{\v r}_{\hat{w}_\nu} \right) - \delta_{\mu\nu}.
\end{equation}
where $\sigma_{\hat{z}_\alpha \hat{w}_\beta}:=\langle(\hat z_\alpha-\bar r_{\hat z_\alpha})(\hat w_\beta-\bar r_{\hat w_\beta})\rangle$ denotes the centred second moment and $\bar r_{\hat{z}_\alpha} = \langle \hat{z}_\alpha\rangle$.
\end{lem}
The proof of the above lemma is given in Appendix~\ref{App:analytics-QFIm}. Lemma~\ref{Lem:QFIm} also enables us to derive the following theorem for the privacy measure:
\begin{thm}[Privacy in Gaussian phase estimation]\label{Thm:privacy-n}
The privacy when estimating the average phase from a pure $M$-mode Gaussian state $\ket{\Psi(\v{\theta})}$, where each local parameter $\theta_\mu$ is encoded via $U_\mu(\theta_\mu) = e^{i\theta_\mu \hat{n}_\mu}$, is given by
\begin{equation}\label{Eq:privacy-general}
    P(\v{Q}, \v v_\text{avg}) \!=\! \frac{2 \displaystyle\sum_{\substack{\mu,\nu \\ z,w \in \{x,p\}}} \left( \sigma_{\hat{z}_\mu \hat{w}_\nu}^2 + 2\, \sigma_{\hat{z}_\mu \hat{w}_\nu} \, \bar{\v{r}}_{\hat{z}_\mu} \bar{\v{r}}_{\hat{w}_\nu} \right)-M}{M\,\left(2 \displaystyle\sum_{\substack{\mu \\ z,w \in \{x,p\}}} \left( \sigma_{\hat{z}_\mu \hat{w}_\mu}^2 + 2\, \sigma_{\hat{z}_\mu \hat{w}_\mu} \, \bar{\v{r}}_{\hat{z}_\mu} \bar{\v{r}}_{\hat{w}_\mu} \right) - M\right)}, 
\end{equation}
where the above summation denotes the double sum over $\mu,\nu$ and $z,w\in\{x,p\}$.
\end{thm}
The proof is in Appendix~\ref{App:analytics-QFIm}.

\section{Main results}\label{Sec:privacy-in-CV}

The framework of Sec.~\ref{Sec:setting-the-scene} reduces the privacy of a distributed phase-sensing protocol to a property of the QFIm. We now take the target direction to be the average-phase direction, $\v u=\v v_{\mathrm{avg}}=\tfrac{1}{\sqrt{M}}(1,\dots,1)^{\ms T}$, and ask what form of privacy can be achieved with a Gaussian probe under unitary local phase encoding. The answer is fundamentally different for two parties and for three or more parties. We start with a no-go theorem (Sec.~\ref{Sec:no-go}), then turn to the exceptional two-party case (Sec.~\ref{Sec:motivating-example}), and finally build a multipartite protocol (Sec.~\ref{Sec:privacy-in-network}).

\subsection{No-go for complete privacy with three or more parties}\label{Sec:no-go}

We first state our main no-go result, which shows that complete privacy of the average phase is impossible for finite-energy Gaussian networks with three or more parties.

\begin{thm}[No-go for complete average-phase privacy]
\label{thm:no-rank-one-average-gaussian-main}
No finite-energy $M$-mode Gaussian probe with $M\ge 3$ undergoing the local phase encoding of Eq.~\eqref{Eq:local-phase-encoding} can achieve complete privacy of the average phase with nonzero sensitivity. More explicitly, if $\v Q$ satisfies
\begin{equation}
\operatorname{range}(\v Q)\subseteq\operatorname{span}\{\v 1\},
\end{equation}
where $\v 1=(1,\dots,1)^{\ms T}$, then $\v Q=\v 0$.
\end{thm}
\begin{sproof}
The full proof is given in Appendix~\ref{App:no-rank-one-average-gaussian}. We sketch the three steps here. \emph{(i)} Unitarity of the encoding implies $\partial_{\v u}\hat\rho = i[\hat G_{\v u},\hat\rho]$ with $\hat G_{\v u} = \sum_\mu u_\mu\hat n_\mu$, so $\v u\in\ker\v Q$ is equivalent to $[\hat G_{\v u},\hat\rho]=0$. Applying this to every $\v u\in\mathcal H_0:=\{\v u:\v 1^{\ms T}\v u=0\}$ forces $\hat\rho$ to be invariant under the entire zero-sum subgroup of local phase rotations. \emph{(ii)} Translating this invariance to phase space using the symplectic action of phase rotations, one shows that the first moments must vanish, every diagonal covariance block must be rotationally symmetric ($\v\sigma_{\mu\mu}=\nu_\mu\mathbbm 1_2$), and—using the existence of a third mode $\ell\neq\mu,\nu$ to construct a zero-sum rotation that affects $\mu$ while leaving $\nu$ fixed—all off-diagonal blocks must vanish ($\v\sigma_{\mu\nu}=0$ for $\mu\neq\nu$). \emph{(iii)} The resulting state $\v\sigma=\bigoplus_\mu\nu_\mu\mathbbm 1_2$, $\bar{\v r}=\v 0$ is invariant under every local phase rotation, not only the zero-sum ones. Hence the QFIm vanishes identically, contradicting any assumption of finite average-phase sensitivity.
\end{sproof}
Importantly, the theorem is specific to the average-phase direction. It does not rule out rank-one Gaussian QFIms aligned with other parameter directions. For example, the finite-energy Gaussian probe $\ket\alpha\!\bra\alpha\otimes\ket 0\!\bra 0^{\otimes(M-1)}$, with $\alpha\neq0$, has $\v Q = 4|\alpha|^2\,\v e_1\v e_1^{\ms T}$, under local phase encoding, and is therefore rank one but supported on the first coordinate. Thus the no-go result arises from requiring complete privacy of the average phase across three or more modes, not from rank one itself.

\subsection{Complete privacy for two parties}\label{Sec:motivating-example}

We turn to the exceptional $M=2$ case identified by Theorem~\ref{thm:no-rank-one-average-gaussian-main} and show that the two-mode squeezed state (TMSS) provides complete average-phase privacy. The example also lets us introduce the main ingredients of interest, such as loss, displacement, measurement strategies, and the role of entanglement. In line with the no-side-information assumption $\textbf{A}_2$, the two parties do not know their local phases prior to inference.

Consider a two-mode squeezed state (TMSS) undergoing local phase shifts $\theta_1$ and $\theta_2$. In this simplest case, one may ask whether it is possible to estimate \emph{only} the average phase. To make the scenario more realistic, we introduce noise prior to the phase encoding, modelled by a beam splitter with transmissivity $\eta$. For simplicity, we assume that both modes experience the same level of loss, even though in experimental implementations this symmetry may not hold. This assumption allows for a cleaner analytical treatment and captures the essential features of the problem; in Appendix~\ref{App:minimal-scenario}, we show how the analysis generalises when the losses differ. Under this symmetric-loss model, it can be shown that the quantum Fisher information matrix of the output state is given by
\begin{equation}\label{Eq:Fisher-motivating-example}
    \v Q(r,\eta) = \frac{\eta^2\sinh^2 2r}{1+2(1-\eta)\eta\sinh^2 r}\begin{pmatrix}
 1 &1 \\
 1 & 1 \\
\end{pmatrix}.
\end{equation}
This matrix is clearly of rank one, capturing the fact that only a single linear combination of the two local phases can be estimated. Diagonalising $\v Q(r,\eta)$, we find that it has a single non-zero eigenvalue $\lambda_{\text{avg}}=2\eta^2\sinh^2 2r/[1+2(1-\eta)\eta\sinh^2 r]$, associated with the eigenvector $\v v_{\text{avg}} = \nicefrac{1}{\sqrt{2}}\,(1,1)^{\ms T}$ corresponding to the average phase $\phi_{\text{avg}} := \nicefrac{1}{2}\,(\theta_1+\theta_2)$. The second eigenvalue vanishes, $\lambda_{\text{rel}} = 0$, with the corresponding eigenvector $\v v_{\text{rel}} = \nicefrac{1}{\sqrt{2}}\,(-1,1)^{\ms{T}}$ linked to the relative phase $\phi_{\text{rel}} :=  \nicefrac{1}{2}\,(\theta_2-\theta_1)$. The vanishing QFI along this orthogonal direction confirms that the relative phase cannot be estimated. Since $\v v_{\text{rel}} \in \operatorname{ker} \v Q$ and both components of $\v v_{\text{rel}}$ are non-zero, according to Definition~\ref{Def:privacy-definition}, both phases $\theta_1$ and $\theta_2$ are private. In fact, since the QFIm is rank one, we have complete privacy, and therefore $P(\v Q,\v v_{\text{avg}}) = 1$.

The two-mode setting falls outside the scope of Theorem~\ref{thm:no-rank-one-average-gaussian-main} because there is no third mode available to force the off-diagonal block $\v\sigma_{12}$ to vanish. The relative-phase symmetry that would otherwise be restrictive reduces to a one-dimensional subgroup, and the TMSS exploits the off-diagonal correlations that this symmetry leaves unconstrained. Although only the average phase remains estimable in the presence of noise, it is important to emphasise that the attainable precision degrades with $\eta$, as the QFIm is reduced by a factor of $\eta^2/[1+2(1-\eta)\eta\sinh^2 r]$.

An interesting case arises when the modes are displaced before encoding the phases. Although this improves the estimation precision (as the quantum Fisher information increases), it compromises privacy. For simplicity, we set $\eta = 1$ and analyse what happens with the spectrum of the QFIm. In contrast with the undisplaced case, the QFIm is no longer rank one and has no zero eigenvalue: both the average-phase direction and directions with relative-phase content can carry finite Fisher information. The two eigenvalues can be written as $\lambda_\pm=f(\alpha_1,\alpha_2,r)\pm g(\alpha_1,\alpha_2,r)$, with eigenvectors that depend on the displacement pattern. We therefore avoid labelling them as purely ``average'' and ``relative'' eigenvalues except in symmetric limiting cases. When $\alpha_1=\alpha_2=0$, we recover the undisplaced TMSS case, where the relative-phase direction is unobservable. As soon as the displacement pattern breaks the photon-number-correlation symmetry, the previously hidden direction acquires finite QFI. Physically, while the undisplaced TMSS is insensitive to opposite local phase shifts because of its perfect photon-number correlations, displacements introduce local phase references and make relative-phase information observable. The loss of privacy by displacing the modes can be quantified by using Eq.~\eqref{Eq:label-privacy-measure}, which gives us
\begin{equation}\label{Eq:example-privacy-alpha}
    P(\v Q, \v v_\text{avg}) = 1- \frac{ \left(\alpha_1^2+\alpha_2^2\right) \cosh 2r+2 \alpha_1 \alpha_2 \sinh 2r}{2 \left(\alpha_1^2+\alpha_2^2\right) \cosh 2r+\sinh^2 2r},
\end{equation}
where, without loss of generality, we assume that $\alpha_1, \alpha_2 \in \mathbbm{R}$. Since $\alpha_1^2+\alpha_2^2\ge |2\alpha_1\alpha_2|$ and $\cosh 2r\ge |\sinh 2r|$, the numerator of the second term in Eq.~\eqref{Eq:example-privacy-alpha} is non-negative, and is strictly positive unless $\alpha_1=\alpha_2=0$. Thus, displacing the modes before encoding the local phases invariably leads to a reduction in the privacy measure. 

We now turn to the question of which measurements the parties are allowed to perform, and what the optimal protocol is for estimating the average phase. If we restrict attention to Gaussian measurements, it can be shown that the optimal strategy within the Gaussian toolbox involves the following scheme: after the local phase encoding, the two modes are interfered on a balanced beam splitter, followed by homodyne measurements in conjugate quadratures rotated by an angle $\phi$ (see Appendix~\hyperref[App-optimal-G-measurement]{C-1}). The optimal angle $\phi_{\text{opt}}$ is given by
\begin{equation}
    \phi_{\text{opt}} = \frac{\theta_1+\theta_2}{2} +\tan^{-1}\qty(\frac{\eta \sinh 2r}{\sqrt{1+4(1-\eta)\eta \sinh^2 r}}).
\end{equation}
However, this dependence of $\phi_{\text{opt}}$ on the unknown average phase poses a practical challenge. This apparent circularity is resolved by standard adaptive metrology: a short pre‑run with heterodyne (or fixed‑angle homodyne) detection provides a coarse estimate of $\phi_{\text{avg}}$. The local oscillators are then iteratively steered towards $\phi_{\text{opt}}$ until the classical Fisher information is maximised. In the lossless case, the resulting Fisher information saturates the quantum benchmark, while under loss, it does not generally attain the QFI. In the regime where phase shifts are small, the average can be neglected, and the optimal angle becomes effectively independent of the unknown parameter, removing the need for adaptive alignment.

Finally, let us consider the role of entanglement for privacy. This naturally leads to a comparison between the previous scenario and the case in which the two-mode squeezed state is replaced by two single-mode squeezed states. Following the same protocol (and assuming $\eta=1$ and $\alpha_1=\alpha_2=0$), it is straightforward to show that the QFIm is $\v Q(r) = 2\sinh^2(2r)\iden_2$, with degenerate spectrum $\lambda_{\text{avg}/\text{rel}}=2\sinh^2(2r)$. Hence $\operatorname{ker}(\v Q)={\v 0}$ and, by Definition~\ref{Def:privacy-definition}, there exists no nonzero vector $\v v$ with $v_j \neq 0$ lying in the kernel. In particular, no component $\theta_j$ is private. This can be seen directly from the privacy measure: for both the average-phase direction $\v v_{\text{avg}}$ and the relative-phase direction $\v v_{\text{rel}}$, we have $P(\v Q, \v v_{\text{avg}}) = P(\v Q, \v v_{\text{rel}}) = \nicefrac{1}{2}$. Thus only half of the available information aligns with the chosen public function, and the remaining half leaks into the orthogonal combination. Equivalently, the scalar QCRB~\eqref{Eq:scalar-QCRB} shows that both combinations are estimated with identical precision. By contrast with the TMSS case, where $\v Q$ is rank-one and concentrates all information along the average-phase direction, two independent single-mode squeezers distribute information uniformly across all phase combinations. This comparison highlights the role of the shared TMSS correlations: the independent squeezers retain sensitivity to the collective target, but their full-rank QFIm provides neither complete nor component privacy.

\subsection{Weak privacy in a quantum network}\label{Sec:privacy-in-network}

\begin{figure}
    \centering
    \includegraphics{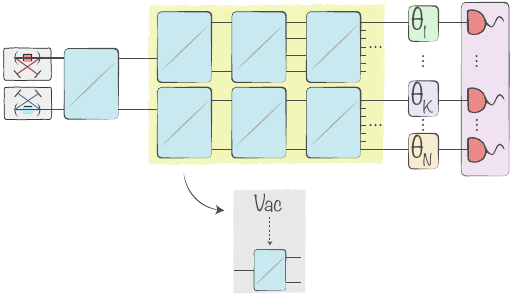}
    \caption{\emph{Protocol}. Two orthogonally squeezed states are generated by optical parametric oscillator (OPOs) and mixed at a central beam splitter to produce a two-mode squeezed state. This is symmetrically distributed via a recursive beam splitter network (yellow area), where each layer mixes modes with vacuum, doubling their number. The resulting state then undergoes local phase  shifts $\{\theta_1, ..., \theta_M\}$, followed by local measurements.
    }
    \label{F-circuit}
\end{figure}

Theorem~\ref{thm:no-rank-one-average-gaussian-main} rules out complete average-phase privacy for $M\ge 3$, since any Gaussian probe whose QFIm has $\operatorname{range}(\v Q)=\operatorname{span}\{\v v_{\mathrm{avg}}\}$ must have $\v Q=\v 0$. For three or more parties, the relevant goal is therefore weaker: every individual phase should remain private, while the normalised QFIm becomes increasingly concentrated along the average-phase direction as the squeezing increases. We now introduce an experimentally feasible protocol with exactly these properties. The protocol proceeds as follows (see Fig.~\hyperref[F-circuit]{\ref{F-circuit}} for a pictorial overview):
\begin{enumerate}
\item A two-mode squeezed state is distributed among $M$ modes using a beam splitter network, with vacuum states injected into the unused input ports.
\item Each output mode undergoes a local phase shift $\theta_i$, where $i \in \{1, \dots, M\}$.
\item Each party performs a local measurement, and the broadcast outcomes are processed centrally to estimate the average phase.
\end{enumerate}

The choice of a two-mode squeezed state is motivated by the two-party example of Sec.~\ref{Sec:motivating-example}. There, privacy comes from a symmetry of the TMSS: the two modes always contain matching photon-number fluctuations, so opposite phase shifts on the two modes cannot be distinguished. The beam-splitter network used below only redistributes the optical fields among more parties. It does not destroy this correlation between the two original input modes. As a result, one collective phase direction remains unobservable after the state is distributed, and this unobservable direction is what makes every individual phase weakly private.  This would not happen if the input resource were a single-mode squeezed vacuum sent through the same balanced beam-splitter network, with vacuum in the unused input ports. In that case, the network would simply spread one squeezed mode over many parties. There would be no second, photon-number-correlated input mode to provide a hidden opposite-phase symmetry. Consequently, all local phase directions would carry some information, the QFIm would have no nontrivial kernel, and no individual phase would be private according to Definition~\ref{Def:privacy-definition}. Thus, weak privacy is not produced by the beam-splitter network alone. It comes from the TMSS symmetry that the network preserves while distributing the state.

As our goal is to analyse the privacy of the protocol, we first construct the quantum Fisher information matrix for the setting at hand. The following lemma gives its explicit form:

\begin{lem}[Beam-splitter-tree QFIm]
\label{lem:tree-QFIM-characterisation}
Consider a two-mode squeezed state split into $2^N$ modes by a balanced $50{:}50$ beam splitter network. 
Encoding local phases $\v\theta=(\theta_1,\dots,\theta_{M})$ via $\hat{U}(\v\theta) =  \bigotimes_{\mu=1}^M e^{i \theta_\mu \hat{n}_\mu}$, the corresponding quantum Fisher information matrix $\v Q$ is independent of $\v\theta$ and can be written in the projector form
\begin{equation}
\v Q = \frac{\cosh 2r - 1}{2^{N-2}}\!\left(\mathbbm{1}_M-\Pi_+-\Pi_0\right) 
+ \frac{\cosh 4r - 1}{2^{N-1}}\,\Pi_+,
\end{equation}
where $\Pi_+ = \v 1\v 1^\top/M$ and $\Pi_0 = \v s\v s^\top/M$ with $\v 1=(1,\dots,1)^\top$ and $\v s=(\underbrace{1,\dots,1}_{\tfrac{M}{2}}, \underbrace{-1,\dots,-1}_{\tfrac{M}{2}})^\top$.
\end{lem}
With an explicit characterisation of the quantum Fisher information matrix in hand, Definition~\ref{Def:privacy-definition} implies that its eigenvalues determine the achievable privacy. The following theorem gives the spectrum of~$\v Q$:

\begin{thm}[Spectrum and eigenmodes of the QFIM]
\label{thm:tree-QFIM-spectrum}
The eigenvalues of the quantum Fisher information matrix $\v Q$ characterised by Lemma~\ref{lem:tree-QFIM-characterisation} are
\begin{equation}\label{Eq:eigenvalues-F}
   \!\!\!\v \lambda(\v Q) 
    \!=\! \left\{
        0,\ 
        \underbrace{\frac{\cosh 2r - 1}{2^{N-2}},\dots,\frac{\cosh 2r - 1}{2^{N-2}}}_{M-2\ \text{times}},\ 
        \frac{\cosh 4r -1}{2^{N-1}}
      \right\}.
\end{equation}
The corresponding orthonormal eigenbasis is given by:
\begin{enumerate}[label=\roman*.]
    \item Zero eigenvalue: $\v v_0 = \frac{\v s}{\sqrt{M}}$.
    \item Degenerate eigenvalue: $\mathcal{W} = \{\v w:\ \v 1^\top\v w = 0,\ \v s^\top\v w = 0\}$.
    \item Largest eigenvalue: $\v v_{\text{avg}} = \frac{\v 1}{\sqrt{M}}$.
\end{enumerate}
\end{thm}

The null direction $\v v_0=\tfrac{\v s}{\sqrt M}$ has a transparent symmetry interpretation. It corresponds to applying the same phase shift to all modes descended from one input arm of the initial TMSS and the opposite phase shift to all modes descended from the other arm. Because the beam-splitter tree is passive, the corresponding generator collapses to $\hat G_0 = \tfrac{1}{\sqrt{M}}(\hat n_A - \hat n_B)$, where $\hat n_A=\sum_{j\in A}\hat n_j$ and $\hat n_B=\sum_{j\in B}\hat n_j$ are the total photon numbers on the two branches. A two-mode squeezed vacuum has perfect photon-number correlations, so $\operatorname{var}(\hat n_A-\hat n_B)=0$. The QFI along $\v v_0$ therefore vanishes by symmetry. Equivalently, the state is invariant under opposite phase rotations of the two TMSS branches, and the beam-splitter network preserves this symmetry. The zero eigenvalue in Eq.~\eqref{Eq:eigenvalues-F} is thus a direct consequence of the input symmetry rather than an accidental cancellation.

The spectrum in Theorem~\ref{thm:tree-QFIM-spectrum} admits a clear metrological interpretation. There are three distinct sectors: one zero eigenvalue (a single direction with vanishing sensitivity), an $(M-2)$-dimensional shot-noise–limited subspace with eigenvalues scaling linearly with the mean photon number, and a unique enhanced direction whose eigenvalue includes a quadratic contribution in the photon number, thus providing a Heisenberg scaling precision benchmark. In other words, among all independent phase combinations, only one collective mode is metrologically special, while all others remain shot-noise limited. A detailed derivation of the photon-number scaling of these eigenvalues is provided in Appendix~\hyperref[App:mean-photon-number]{D-4}

Combining Lemma~\ref{lem:tree-QFIM-characterisation} with Theorem~\ref{thm:tree-QFIM-spectrum}, we can now determine the privacy properties of this protocol. This is done via the following corollary:

\begin{cor}[Private components] \label{cor:privacy-weak-not-strong}
For a TMSS split into $M=2^N$ modes via a balanced $50{:}50$ beam splitter network, with local phases encoded through phase shifts:
\begin{enumerate}[label=(\roman*),itemsep=2pt,topsep=2pt]
\item The kernel of $\v Q$ is one-dimensional, with $\v v_0\in\ker\v Q$, so there exists a single unobservable direction and the scheme exhibits weak privacy.
\item Since every entry of $\v s$ is nonzero, each parameter $\theta_j$ is individually private.
\item The QFIm is not rank-one: there are $M-2$ degenerate eigenvalues $(\cosh 2r-1)/2^{N-2}$ associated with $\mathcal W$. Hence complete privacy does not hold.
\end{enumerate}
\end{cor}

Item \emph{(iii)} is also a structural consequence of Theorem~\ref{thm:no-rank-one-average-gaussian-main}: complete privacy is impossible for $M\ge 3$ Gaussian probes with unitary phase encoding, so any concrete protocol can at best achieve the weak privacy described in items \emph{(i)}--\emph{(ii)}. The proofs of these three results are given in Appendix~\ref{App:covariance-two-mode-squeezed-state-QFIM}. We also provide an efficient \texttt{Mathematica} notebook that constructs the covariance matrix of the beam splitter tree, the quantum Fisher information matrix, and its eigenvalues for an arbitrary number of nodes in the network--along with many other results concerning privacy in CV systems~\cite{adeoliveirajunior2025d}. The above results assume a specific resource state, namely a two-mode squeezed state, and a specific protocol, consisting of a beam-splitter network followed by local phase encoding. For a generic pure state produced by a Gaussian protocol and subjected to local phase shifts, the QFIm can be computed using Lemma~\ref{Lem:QFIm}. In the case of interest--where $|\psi_0\rangle$ is a two-mode squeezed state distributed through a beam splitter network--Lemma~\ref{Lem:QFIm} recovers Lemma~\ref{lem:tree-QFIM-characterisation} (and vice versa). 

We summarise the message of this section. For $M=2$ the TMSS achieves complete privacy, and Theorem~\ref{thm:no-rank-one-average-gaussian-main} closes the door on this for any $M\ge 3$. When the TMSS is distributed among more parties via the beam-splitter tree, every individual phase remains private (Corollary~\ref{cor:privacy-weak-not-strong}), but certain orthogonal combinations leak; the leaked information is, however, insufficient to reconstruct any individual phase under the no-side-information assumption $\textbf{A}_2$. In linear-algebra terms, the encoded data leave an underdetermined affine family of compatible parameter vectors $\v\theta+\ker\v Q$, so no individual component is identifiable from the public record alone without additional side information. We now turn to a quantitative analysis of these results.

\section{Privacy performance and imperfections}\label{Sec:privacy-optimality}

\begin{figure}
    \centering
    \includegraphics{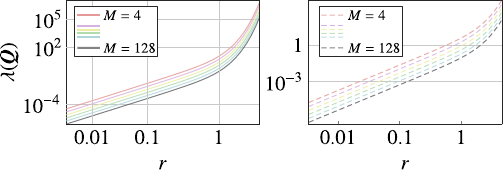}
    \caption{\emph{Quantum Fisher information spectrum}. Log–log plot of the eigenvalues of the quantum Fisher information matrix $\v \lambda(\v Q)$ as a function of the squeezing parameter $r$, for different numbers of modes $M=2^N$. The largest eigenvalue is shown with solid lines, while the smallest nonzero eigenvalue is shown with dashed lines. As $M$ increases, the spectral gap between the two eigenvalues widens, highlighting how the distribution of information across phase directions depends on both squeezing and network size. The total number of photons (and hence the total metrological resource) is conserved; splitting the state among more modes spreads these photons more thinly across the network, so each local phase shift affects fewer photons and the sensitivity per parameter direction decreases.
    }
    \label{F-privacy-eig}
\end{figure}

For $M\ge 3$, complete average-phase privacy cannot be achieved by finite-energy Gaussian probes (Theorem~\ref{thm:no-rank-one-average-gaussian-main}). However, privacy at the level of individual phases remains guaranteed for the TMSS-tree protocol (Corollary~\ref{cor:privacy-weak-not-strong}). We now analyse the privacy performance of this protocol under displacement and loss, and compare it with selected alternative Gaussian resources.

In Fig.~\hyperref[F-privacy-eig]{\ref{F-privacy-eig}} we plot the two non-zero eigenvalues of the QFIm [Eq.\eqref{Eq:eigenvalues-F}] as functions of the squeezing parameter on a log–log scale, for different values of $M$ (colours ranging from red to black). The left and right panels show the largest and smallest non-zero eigenvalues, respectively. As the squeezing increases, the dominant eigenvalue grows as $\cosh 4r$ while the subdominant eigenvalue grows as $\cosh 2r$, both diverge in absolute terms, but the ratio between them grows as $\cosh 4r/\cosh 2r\sim e^{2r}$ at large $r$. Consequently, the fraction of the total Fisher information lying outside the average-phase direction vanishes asymptotically, and the normalised spectrum approaches the rank-one limit, even though no eigenvalue is in fact suppressed. With a fixed total mean photon number, increasing the number of modes $M$ spreads the photons across more parameters; as a result, each local phase receives fewer photons and the shot-noise–limited sensitivity per direction decreases.

Applying Theorem~\ref{Thm:privacy-n}, one can get a closed-form expression for the privacy as a function of the number of parties ($M=2^N)$ in the average phase estimation, namely we have that
\begin{equation}\label{Eq:privacy-scheme}
    P(\v Q, \v v_{\text{avg}}) = 1 - \frac{2^N - 2}{[(2^N - 1) + \cosh 2r]} \quad \text{for} \quad r>0,
\end{equation}
Reaching $P=1$ in the multi-party setting would require the unphysical infinite-squeezing limit, consistent with Theorem~\ref{thm:no-rank-one-average-gaussian-main}. At first sight, this behaviour may seem paradoxical. Although the privacy measure falls approximately as $1/M$, the QFIm always contains a strictly zero eigenvalue, ensuring that at least one linear combination of local phases remains completely hidden. This reflects the structure of the beam splitter network. Each added layer doubles the number of output modes and enlarges the phase subspace orthogonal to the protected direction. Besides the fully hidden branch-imbalance vector $ s=(\underbrace{1,\ldots,1}_{\tfrac{M}{2}}, \underbrace{-1,\ldots,-1}_{\tfrac{M}{2}})^\top$, defined in Lemma~\ref{lem:tree-QFIM-characterisation}, the QFIm has an $(M-2)$-dimensional degenerate eigenspace $ W=\{w\in\mathbb R^M:\mathbf 1^\top w=0,\ s^\top w=0\}. $ Writing $A=\{1,\ldots,\tfrac{M}{2}\}$ and $B=\{\tfrac{M}{2}+1,\ldots,M\}$ for the two branches of the tree, the conditions $\mathbf 1^\top w=0$ and $s^\top w=0$ are equivalent to $ \sum_{j\in A}w_j=0, \qquad \sum_{j\in B}w_j=0 . $ Thus $W$ consists of phase contrasts within each branch, such as $e_1-e_2$ or $e_{\tfrac{M}{2}+1}-e_{\tfrac{M}{2}+2}$, rather than contrasts between the two branches. All directions in $W$ have the same finite eigenvalue $(\cosh 2r-1)/2^{N-2}$. With $r$ held fixed, the total Fisher information must be distributed across more directions, so the share associated with the average phase diminishes. Henceforth we restrict attention to the four-mode instance $M=4$, the smallest network that already displays all qualitative features (a single fully hidden direction, plus degenerate orthogonal sensing directions) and is experimentally realisable. Using Eq.~\eqref{Eq:privacy-scheme}, $P_4(r):=1-\frac{2}{3+\cosh 2r}$.
\begin{figure}[t]
    \centering
    \includegraphics{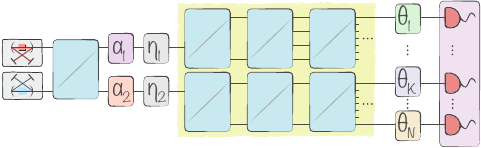}
    \caption{\emph{Displacement \& loss}. The two-mode squeezed state is first displaced by amounts $\alpha_1$ and $\alpha_2$, and then subjected to a loss channel modelled by beam splitters with transmissivities $\eta_1$ and $\eta_2$, respectively. Subsequently, the modes undergo the beam splitter network, after which local phases are encoded.
    }
    \label{F-displacement-noise}
\end{figure}

We now consider a slightly more general scenario by introducing displacement operations on both modes before they are sent through the beam splitter network (see Fig.~\ref{F-displacement-noise}). First, as in the two-mode case, displacements remove the strictly zero eigenvalue of the QFIm, so that no parameter direction remains completely hidden. However, the smallest eigenvalue can remain very small ($\epsilon$-close to zero). Second, by Theorem~\ref{Thm:privacy-n}, the privacy measure when the TMSS is split into four modes is $P(\v Q,\v v)=P_4(r)-f(r,\alpha_1,\alpha_2)$, where $f(r,\alpha_1,\alpha_2)\ge 0$ quantifies the reduction in privacy due to the displacements.

\begin{figure}[t]
    \centering
    \includegraphics{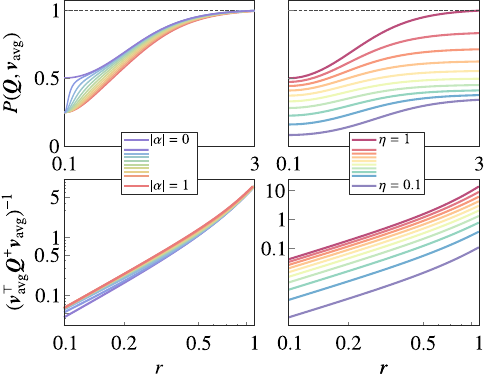}
    \caption{\emph{Privacy under displacement \& loss}. Top panels: privacy for different values of $|\alpha|=|\alpha_2-\alpha_1|$ in the absence of loss (left), and privacy in the presence of a loss channel with efficiency $\eta$ for fixed $|\alpha|=0$ (right). Bottom panels: effective scalar QFI $(\v v_{\rm avg}^{\ms T}\v Q^{+}\v v_{\rm avg})^{-1}$ for the average-phase target, which is in $\operatorname{range}(\v Q)$ throughout the plotted regimes, so the pseudoinverse expression is well-defined.}
    \label{F-privacy-r-eta}
\end{figure}

In the first panel of Fig.~\ref{F-privacy-r-eta} we examine privacy in the absence of loss ($\eta=1$) while displacing the TMSS by $|\alpha|=|\alpha_2-\alpha_1|$ in phase space. In the plotted curves we restrict to the displacement path specified in the numerical calculation, so that $|\alpha|$ denotes the chosen relative displacement amplitude rather than a complete parametrisation of the two-dimensional displacement space. The privacy measure decreases with increasing $|\alpha|$, although it asymptotically approaches one as $r\to\infty$. Since displacement is a resource for parameter estimation, the reduction in privacy arises from information leakage at the level of the encoded state: displacements break the symmetry in phase space and individual phases are imprinted in the first moments. Operationally, any unintended displacement can be diagnosed by estimating the first moments of each mode (which ideally vanish in the undisplaced protocol).

Correspondingly, the scalar Fisher information $(\v v^{\ms T}\v Q^{+}\v v)^{-1}$ (bottom panel of Fig.~\ref{F-privacy-r-eta}) increases with $|\alpha|$. Importantly, values of $P$ close to $1$ should not by themselves be read as ``more secrecy''. When overall sensitivity is depleted, $P\to 1$ can simply indicate that nearly all information has been erased rather than genuinely hidden. In the second panel we replace displacement with a loss channel of transmissivity $\eta$. Loss degrades the privacy measure, but this reduction does not arise from information leakage---the zero eigenvalue of the QFIm persists. Rather, it reduces the fraction of the total Fisher information aligned with the target. A possible way to improve privacy in the presence of loss is to allow finite displacement and optimise over $|\alpha|$, increasing precision at the cost of partially sacrificing privacy.

\begin{figure}[t]
    \centering
    \includegraphics{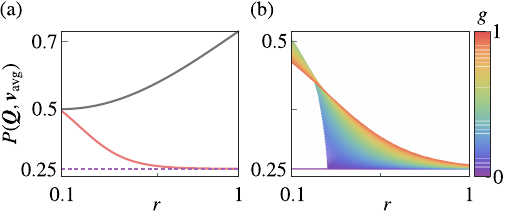}
    \caption{\emph{Privacy comparison}. (a) Privacy $P(\v Q,\v v_{\text{avg}})$ as a function of $r$ for three states: TMSS distributed via a symmetric beam splitter network (black), a linear cluster state on four parties via sequential $\operatorname{CZ}$ operations on $\{(1,2),(2,3),(3,4)\}$ at the optimal coupling $g_{\operatorname{opt}}\approx 0.88$ (red), and a product of single-mode squeezed states (dashed). (b) Privacy of the cluster state for various $g$ versus $r$.}
    \label{F-privacy-differentstates}
\end{figure}

Having characterised the impact of imperfections in the TMSS-based protocol, we now ask whether alternative Gaussian states might offer improved or comparable privacy. In particular, we explore whether continuous-variable analogues of graph states---linear cluster states generated via $\operatorname{CZ}$ operations---can support private estimation of the average phase. We compare the privacy of cluster states with that of both the TMSS-split network and separable squeezed product states. We benchmark the privacy of different Gaussian states under average phase estimation (see Fig.~\ref{F-privacy-differentstates}). A TMSS distributed symmetrically via beam splitters concentrates an increasing fraction of the Fisher information along the target direction, leading to $P\to 1$ in the large-squeezing limit even though the off-target eigenvalues do not vanish absolutely. In contrast, cluster states do not support private estimation: their privacy measure asymptotically converges to the separable-state baseline $P=\tfrac14$. Increasing the $\operatorname{CZ}$ coupling strength does not improve privacy and, in fact, leads to greater information leakage about individual parameters, suggesting a fundamental incompatibility between the position-position correlations introduced by $\operatorname{CZ}$ operations and the number-based phase encoding used here. Whether the TMSS-tree protocol is optimal among all Gaussian probes under the no-go constraint of Theorem~\ref{thm:no-rank-one-average-gaussian-main} remains an open question. For the specific protocol considered here, however, we provide both analytical and numerical evidence that it is near-optimal within the family of Gaussian resources we tested.

\section{Conclusion and outlook}

We have introduced and analysed an information-theoretic characterisation of privacy for continuous-variable sensor networks. Our framework is based on unobservable directions, namely vectors lying in the kernel of the QFIm, and provides an operational way to assess privacy at the level of individual components under the prescribed-protocol and no-side-information assumptions $\textbf{A}_1$ to $\textbf{A}_4$. The central structural result is Theorem~\ref{thm:no-rank-one-average-gaussian-main}, which states that, under unitary local phase encoding, complete privacy of the average phase is impossible at finite energy for three or more modes. The two-mode squeezed state is the exceptional Gaussian case that achieves complete privacy. For the TMSS distributed over $M=2^N$ modes, we derived a closed-form projector expression for the QFIm in Lemma~\ref{lem:tree-QFIM-characterisation}, obtained its full spectrum and eigenmodes in Theorem~\ref{thm:tree-QFIM-spectrum}, and translated these results into operational privacy statements in Corollary~\ref{cor:privacy-weak-not-strong}. The protocol has a single unobservable direction and makes every local phase individually private, although complete privacy is not attained at finite squeezing. We also obtained general closed-form expressions for the QFIm of arbitrary pure multimode Gaussian states undergoing local phase shifts in Lemma~\ref{Lem:QFIm} and for the privacy measure $P(\v Q,\v v_{\mathrm{avg}})$ in Theorem~\ref{Thm:privacy-n}.

We further analysed two practically relevant scenarios. Displacements enhance estimation precision by imprinting phases into the first moments but simultaneously remove the zero eigenvalue of the QFIm, destroying privacy at the component level. Optical loss preserves the zero eigenvalue, so no individual phase becomes identifiable, but it reduces the fraction of the total Fisher information aligned with the target direction, lowering the privacy measure. A comparative study revealed that alternative Gaussian resources, such as four-mode cluster states or independent single-mode squeezers, perform worse than the TMSS-tree protocol within the family we tested. A natural question under the no-go constraint of Theorem~\ref{thm:no-rank-one-average-gaussian-main} is to identify the optimal protocol among experimentally friendly Gaussian resources.

The privacy guarantees of this work rely on the access-controlled encoding assumption $\textbf{A}_2$ and on the adversarial model $\textbf{A}_4$: each party can use only the public broadcast and the quantum systems in its possession to infer information about the local phases, has no additional query access to any encoding channel, and does not know any raw value $\theta_\mu$ directly. If adversaries are instead allowed to know their own raw phases, or to interrogate their encoding channels with auxiliary probes, the privacy criterion used here is no longer sufficient. Indeed, for the TMSS-tree protocol the QFIm has a one-dimensional kernel generated by a vector with nonzero support on every component. Fixing even one component of $\v\theta$ then removes the remaining ambiguity along this kernel direction and can reveal the other components.  A natural strengthening, which we leave for future work, is conditional privacy. In this stronger sense, $\theta_j$ would remain private against an adversarial subset $S\not\ni j$ if there exists a kernel direction $\v w\in\ker\v Q$ such that $w_j\neq 0$ while $w_i=0$ for every $i\in S$. Designing continuous-variable protocols whose QFIm kernels satisfy this stronger condition is an open problem.
While our analysis assumes an honest setting, this is an idealisation that may not hold in practical network deployments. Another extension is therefore to incorporate a certification layer, enabling participants to verify in situ that the distributed resource is sufficiently close to the intended state. In the discrete-variable regime, Shettell \emph{et al.}~\cite{shettell2022} have shown how stabiliser-based tests can certify GHZ-type states. More recently, Descamps and Markham~\cite{descamps2024} extended this result to continuous-variable graph states, developing noise-tolerant verification routines. An open problem is to develop a certification protocol tailored to our specific scheme.

Finally, our scheme can be feasibly implemented using a source of two-mode squeezed states, an array of beam splitters, and homodyne detectors. Two single-mode squeezed states can be generated via parametric down-conversion~\cite{wu1986generation} or four-wave mixing~\cite{slusher1985observation} and interfered on a balanced beam splitter to produce a two-mode squeezed vacuum. This TMSS is then distributed to four parties by an additional beam-splitter network. Each party's local phase is imprinted by a channel that the party does not control, after which homodyne measurements of appropriately chosen quadratures on each mode are performed to reconstruct the covariance matrix of the state. The state preparation, distribution, and detection stages thus rely only on well-established optical components. We emphasise, however, that the Heisenberg-scaling precision discussed in this work is a benchmark defined under unrestricted measurements: constructing a strictly local measurement strategy that attains it for $M\ge3$, or quantifying the unavoidable gap, remains an important open step towards a fully distributed implementation.

\begin{acknowledgments}
AOJ would like to thank Mohammad Mehboudi for the helpful discussions. The authors acknowledge financial support from the  EU Horizon Europe (QSNP, grant no. 101114043 \& CLUSTEC, grant no. 101080173 \& ClusterQ, grant no. 101055224), NNF project CBQS (NNF 24SA0088433), the Danish National Research Foundation grant bigQ (DNRF 142). MT acknowledges JST Moonshot R\&D (JPMJMS226C and JPMJMS2061), JST CRONOS (JPMJCS24N6), and JST ASPIRE (JPMJAP2427). SWM and DM acknowledge support from the ANR project EQUINE (ANR-23-QUAC-0001).
\end{acknowledgments}

\bibliography{References}

\clearpage 
\appendix
\onecolumngrid

\section{Unobservability \& privacy}\label{App:privacy-def}

In this appendix we prove Proposition~\ref{Prop:privacy-lemma}. We first present a lemma giving equivalent characterisations of unobservability at regular points of the quantum statistical model; the unitary phase-encoding family used in the main text falls in this class, as discussed below Definition~\ref{Def:privacy-definition}.
\begin{lem}[Equivalent criteria for unobservability at regular points]\label{Lem:equivalence}
Suppose the multiparameter statistical model is regular at $\v\theta$, in the sense that $\hat\rho(\v\theta)$ has locally constant rank and the SLDs are well defined on $\operatorname{supp}(\hat\rho)$. Let $\v Q$ be the QFIm with entries $Q_{\mu\nu}=\tfrac12\tr[\hat\rho\{\hat L_{\theta_\mu},\hat L_{\theta_\nu}\}]$, where $\hat L_{\theta_\mu}$ are the SLDs. Then, for any real vector $\v u\in\mathbb R^M$, the following are equivalent:
\begin{enumerate}
\item[(i)] $\partial_{\v u}\hat\rho=0$,
\item[(ii)] $\hat L_{\v u}:=\sum_\mu u_\mu\hat L_{\theta_\mu}=0$ on $\operatorname{supp}(\hat\rho)$,
\item[(iii)] $\v u\in\ker\v Q$.
\end{enumerate}
\end{lem}
\begin{proof}
Linearity of the SLD equation gives $2\partial_u\hat\rho=\hat\rho\hat L_u+\hat L_u\hat\rho, \qquad \hat L_u:=\sum_\mu u_\mu \hat L_{\theta_\mu}. $ Moreover, $\v u^\top \v Q\v u=\tr(\hat\rho \hat L_u^2)\ge0. $ Thus $\v u\in\ker \v Q$ if and only if $\tr(\hat\rho \hat L_u^2)=0$, which is equivalent to $\hat L_u=0$ on $\operatorname{supp}(\hat\rho)$. The SLD equation then gives $\partial_u\hat\rho=0$ on the support. At regular points the rank of $\hat\rho(\theta)$ is locally constant, so the SLD equation represents the full derivative without additional singular terms from eigenvalues emerging out of zero. Hence $\partial_u\hat\rho=0$.  Conversely, if $\partial_u\hat\rho=0$, then the SLD equation gives $ \hat\rho\hat L_u+\hat L_u\hat\rho=0. $ In an eigenbasis of $\hat\rho$, with eigenvalues $p_a$, this reads $(p_a+p_b)(L_u)_{ab}=0. $ Whenever at least one of $a,b$ belongs to the support of $\hat\rho$, one has $p_a+p_b>0$, and therefore $(L_u)_{ab}=0$. Hence $\hat L_u=0$ on $\operatorname{supp}(\hat\rho)$, and $ \v u^\top \v Q\v u=\tr(\hat\rho \hat L_u^2)=0. $ Since $Q\succeq0$, this is equivalent to $\v u\in\ker \v Q$.
\end{proof}

\section{Privacy measure in GHZ, W, and product states}\label{App-simple-example}

To illustrate the notion of privacy, consider a simple phase-estimation protocol. Specifically, three parties sharing a state $\ket{\Psi}$ each locally encode a phase $\theta_j$ by applying the unitary operation $U(\theta_j) = \ketbra{0}{0} + e^{i\theta_j}\ketbra{1}{1}$. Our goal is to assess privacy when the parties attempt to estimate the average phase $\theta_{\text{avg}} = (\theta_1 + \theta_2 + \theta_3)/3$. This can be achieved by computing $P(\v Q, \v v_{\text{avg}})$, where in this case $\v v_{\text{avg}} = \frac{1}{\sqrt{3}}(1,1,1)$. Consequently, the problem reduces to calculating the quantum Fisher information matrix $\v Q$ with respect to the average phase. As we are dealing with pure states, the QFIm is simply given by
\begin{equation}
    \v{Q}_{\mu \nu}(\v \theta) = 4\text{Re}\qty[\braket{\partial_{\theta_{\mu}} \Psi(\v \theta)}{\partial_{\theta_{\nu}} \Psi(\v \theta)}-\braket{\partial_{\theta_\mu} \Psi(\v \theta)}{\Psi(\v \theta)}\braket{\Psi(\v \theta)}{\partial_{\theta_\nu} \Psi(\v \theta)}].
\end{equation}
To highlight the non-trivial nature of $P(\v Q, \v v_{\text{avg}})$, we consider two different state preparations undergoing the local encoding. Namely,
\begin{align}
    \ket{\text{GHZ}} = \frac{1}{\sqrt{2}}(\ket{000}+\ket{111}) \xrightarrow[]{U(\v \theta)} \ket{\Phi_\text{GHZ}(\v \theta)}&= \frac{1}{\sqrt{2}}(\ket{000}+e^{i(\theta_1+\theta_2+\theta_3)}\ket{111}). \\
    \ket{\text{W}}= \frac{1}{\sqrt{3}}(\ket{001}+\ket{010}+\ket{100}) \xrightarrow[]{U(\v \theta)} \ket{\Phi_+(\v \theta)} &= \frac{1}{\sqrt{3}}(e^{i\theta_3}\ket{001}+e^{i\theta_2}\ket{010}+e^{i\theta_1}\ket{100}).
\end{align}
These two cases illustrate opposite extremes: the GHZ state allows complete privacy for the average phase, whereas the W state is blind to the average phase. To see this, we compute the quantum Fisher information for each state:
\begin{equation}
    \v Q_{\text{GHZ}} = \begin{pmatrix}
        1 & 1& 1 \\ 
        1 & 1& 1\\
        1 & 1& 1
    \end{pmatrix} \quad , \quad \v Q_{\text{W}} = \frac{4}{9}\begin{pmatrix}
        2 & -1& -1 \\ 
        -1 & 2& -1\\
        -1 & -1& 2
    \end{pmatrix}.
\end{equation}
From their quantum Fisher information, we observe that $\v Q_{\rm GHZ}$ is a rank-one matrix proportional to $\v v_{\rm avg}\v v_{\rm avg}^T$. By contrast, $\v Q_W$ has rank two and $ \ker \v Q_W=\operatorname{span}\{(1,1,1)^\top\}. $ Thus the average-phase direction $v_{\rm avg}=\tfrac{1}{\sqrt{3}}(1,1,1)^\top$ lies in the kernel of $\v Q_W$, so the average phase is completely inaccessible from this state. Nevertheless, since the null vector has nonzero support on every component, each individual phase is private in the sense of Definition~\ref{Def:privacy-definition}. We therefore have $P(\v Q_{\rm GHZ},\v v_{\rm avg})=1$ and $P(\v Q_W,\v v_{\rm avg})=0$.

As a final case, we now consider the scenario in which the parties share product states and each encodes the local phase, leading to
\begin{equation}
    \ket{+_3} = \qty[\frac{1}{\sqrt{2}}(\ket{0}+\ket{1})]^{\otimes 3} \xrightarrow[]{U(\v \theta)} \ket{\Phi_+(\v \theta)} = \bigotimes_{j=1}^3\frac{1}{\sqrt{2}}\qty(\ket{0}+e^{i\theta_j}\ket{1}).
\end{equation}
In this case, the quantum Fisher information is simply $\v Q_+ = \iden_3$. This implies that each phase $\theta_j$ can be independently and optimally estimated. Nothing is hidden--all functions are accessible, as the QFIm is full rank and every direction in parameter space is estimable. Even though the product state leaks all information (i.e., there is no privacy in the conventional sense), the privacy measure $P(\v Q, \v{v})$ is still non-zero--specifically, $P(\v Q, \v v_{\text{avg}}) = \frac{1}{3}$. The fact that $P$ is not zero reflects that the desired function is indeed accessible, but information about other functions is also present.

\section{Minimal scenario for privacy}\label{App:minimal-scenario}

Consider a scenario in which two parties share a two-mode squeezed state, characterised by the covariance matrix
\begin{equation}\label{Eq:app:tms-vac}
 \v \sigma = \frac{1}{2} \begin{pmatrix}
 \cosh 2r & 0 & -\sinh 2r & 0 \\
 0 & \cosh 2r & 0 & \sinh 2r \\
 -\sinh 2r & 0 & \cosh 2r & 0 \\
 0 & \sinh 2r & 0 & \cosh 2r \\
\end{pmatrix},
\end{equation}
with a first-moment vector given by $\bar{\v r} = \v 0$. To maintain generality, one can assume an imperfect state preparation by sending the two-mode squeezed state through a Gaussian loss channel. This can be modelled by coupling both modes to a vacuum mode via a beam splitter with transmissivity $\eta$. After this transformation, the covariance matrix of the output state is given by:
\begin{equation}\label{Eq:covariance-matrix-2-mode-loss}
 \v \sigma_{\text{loss}} = \frac{1}{2} \begin{pmatrix}
 1+2\eta\sinh^2 r & 0 & -\eta\sinh 2r & 0 \\
 0 & 1+2\eta\sinh^2 r & 0 & \eta\sinh 2r \\
 -\eta\sinh 2r & 0 & 1+2\eta\sinh^2 r & 0 \\
 0 & \eta\sinh 2r & 0 &1+2\eta\sinh^2 r \\
\end{pmatrix},
\end{equation}
Assuming that both parties encode a phase $\theta_j$ via a phase shift, the encoded covariance matrix takes the following form:
\begin{equation}
 \v \sigma_{\v \theta} = \frac{1}{2} \begin{pmatrix}
 1+2\eta\sinh^2 r & 0 & -\eta\cos(\theta_1+\theta_2)\sinh 2r & -\eta\sin(\theta_1+\theta_2)\sinh 2r \\
 0 & 1+2\eta\sinh^2 r & -\eta\sin(\theta_1+\theta_2)\sinh 2r & \eta\cos(\theta_1+\theta_2)\sinh 2r \\
 -\eta\cos(\theta_1+\theta_2)\sinh 2r & -\eta\sin(\theta_1+\theta_2)\sinh 2r & 1+2\eta\sinh^2 r & 0 \\
-\eta\sin(\theta_1+\theta_2)\sinh 2r & \eta\cos(\theta_1+\theta_2)\sinh 2r & 0 &1+2\eta\sinh^2 r \\
\end{pmatrix},
\end{equation}
The symmetric logarithmic derivatives (SLDs) can be directly obtained when $\eta = 1$. However, for an arbitrary $\eta$, the calculation is not as straightforward. Nonetheless, we can construct an ansatz for the SLD. More precisely, both symmetric logarithmic derivatives are identical and given by:
\begin{equation}\label{Eq:ansatz-L}
\mathcal{L}_{\v{\theta}_{1,2}}=\frac{\sinh 2r}{2+4\eta(1-\eta)\sinh^2 r}\begin{pmatrix}
 0 & 0 &  \sin (\theta_1+\theta_2)  & -\cos (\theta_1+\theta_2)  \\
 0 & 0 & -\cos (\theta_1+\theta_2)  & - \sin (\theta_1+\theta_2)  \\
 \sin (\theta_1+\theta_2)  &  -\cos (\theta_1+\theta_2) & 0 & 0 \\
  -\cos (\theta_1+\theta_2)  &-\sin (\theta_1+\theta_2)   & 0 & 0 \\
\end{pmatrix}.
\end{equation}
Observe that Eq.~\eqref{Eq:ansatz-L} satisfies $\mathcal{D}_{\v \sigma}(\mathcal{L}_{\v{\theta}_{1,2}}) = 2\partial_{\theta_1,\theta_2} \v \sigma$.

Thus, it is straightforward to show that the quantum Fisher information matrix in this case is rank-1 and given by
\begin{equation}
    \v Q = \frac{\eta^2\sinh^2 2r}{1+2(1-\eta)\eta\sinh^2 r}\begin{pmatrix}
 1 &1 \\
 1 & 1 \\
\end{pmatrix}.
\end{equation}
This implies that $\v P (\v Q,\v v_{\text{avg}}) = 1$. In the case where there are no losses $\eta = 1$ and $\v Q \propto \sinh^2 2r $ as expected.

\subsection{Optimal measurement}\label{App-optimal-G-measurement}

We now show that the optimal measurement for estimating the average phase $\theta_{\text{avg}} := (\theta_1 + \theta_2)/2$ in our two-mode setup is a Gaussian measurement--specifically, an adaptive scheme in which the two-mode squeezed state, after local phase encoding, is passed through a 50:50 beam splitter, followed by homodyne detection at an optimal angle $\phi_{\text{opt}}$. We begin with the state after the phase shifts, whose covariance matrix is given by Eq.~\eqref{Eq:covariance-matrix-2-mode-loss}. The two encoded modes are interfered on a single balanced
$50{:}50$ beam splitter. This transformation renders the output state’s covariance matrix block-diagonal:
\begin{equation}
    \v \sigma_{\v \theta, \text{bs}} = 
\frac{1}{2}\begin{pmatrix} \v A & \v 0 \\ \v 0 & \v B
\end{pmatrix}
\end{equation}
where $\v A$ and $\v B$ are matrices given by
\begin{align}
\v A &= \frac{1}{2}
\setlength{\arraycolsep}{-0.5pt} 
\begin{pmatrix}
1 - \eta(2\sinh^2 r - \sinh 2r \cos\theta) & -\eta \sinh 2r \sin\theta \\
-\eta \sinh 2r \sin\theta & 1 + \eta(2\sinh^2 r + \sinh 2r \cos\theta)
\end{pmatrix} \\ \v B&= \frac{1}{2}
\setlength{\arraycolsep}{-0.5pt} 
\begin{pmatrix}
1 - \eta(2\sinh^2 r + \sinh 2r \cos\theta) & \eta \sinh 2r \sin\theta \\
\eta \sinh 2r \sin\theta & 1 - \eta(\sinh 2r \cos\theta-2\sinh^2 r)
\end{pmatrix}
\end{align}
The symmetry and block-diagonal form of the covariance matrix allows us to analyse each output port independently, simplifying the analysis of measurement strategies. We now turn to the measurement stage, where homodyne detection is performed on each output port of the beam splitter. Port 1 is measured in a rotated $\hat{x}_\phi$ quadrature, while port 2 is measured in the conjugate $\hat{p}_\phi$ quadrature. Specifically, the measurement covariance matrices are taken in the limiting form:
\begin{equation}
    \v \sigma^{(1)}_{\ms M} = R(\phi)\operatorname{diag}\qty(\frac{1}{r_1}, r_1)R(\phi)^{\ms T} \quad \text{and} \quad   \v \sigma^{(2)}_{\ms M} = R(\phi)\operatorname{diag}\qty(r_2, \frac{1}{r_2})R(\phi)^{\ms T} 
\end{equation}
where $R(\phi)$ is a rotation matrix with $r_1$ and $r_2$ being squeezing parameters. In the limit $r_1 \to \infty$ and $r_2\to\infty$, these measurements become projective: Party 1 effectively measures $\hat{x}_\phi$, and Party 2, $\hat{p}_\phi$, with infinite precision along those directions and maximal uncertainty in their conjugates.

The classical Fisher information can be expressed in terms of the first and second moments as follows~\cite{Cenni2022thermometryof}
\begin{equation}\label{Eq:classical-Fisher}
\mathcal{F}_C(\theta;\v r_\theta,\v \sigma_\theta;\v \sigma_{\ms M}) = \frac{1}{2}\tr{\qty[\qty(\v \sigma_\theta + \v \sigma_{\ms M})^{-1} \qty(\frac{\partial \v \sigma_{\theta}}{\partial \theta})]^2} + \qty(\frac{\partial \v r_{\theta}}{\partial \theta})^{\top} \qty(\v \sigma_\theta + \v \sigma_{\ms M})^{-1} \qty(\frac{\partial \v r_{\theta}}{\partial \theta}). 
\end{equation} 
The optimal measurement angle is obtained by maximising $\mathcal{F}_C(\phi)$ with respect to $\phi$. This yields:
\begin{equation}
    \phi_{\text{opt}} = \frac{\theta_1+\theta_2}{2} +\tan^{-1}\qty(\frac{\eta \sinh 2r}{\sqrt{1+4(1-\eta)\eta \sinh^2 r}}).
\end{equation}
Finally, plugging $\phi_{\text{opt}}$ into the classical FI yields:

\begin{equation}
    \mathcal{F}_C(r) = \frac{2\eta^2 \sinh^2 2r}{1+4(1-\eta)\eta \sinh^2 r}.
\end{equation}

\begin{figure*}
    \centering
    \includegraphics{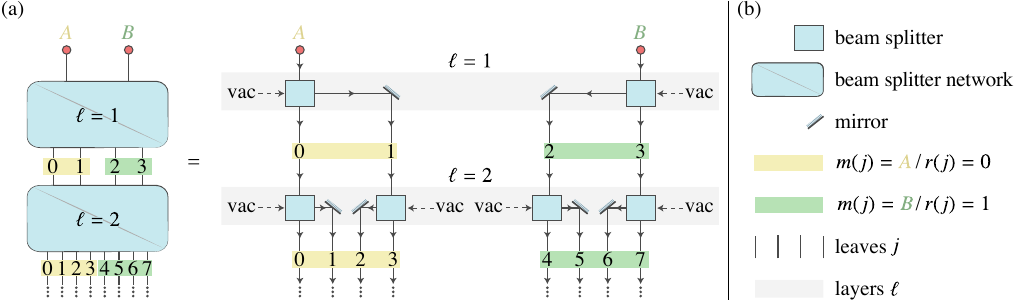}
    \caption{\emph{Beam-splitter tree}. (a) A two-mode squeezed state $(A,B)$ is split into $M=2^N$ modes ($N \ge 2$) by a balanced beam–splitter tree. At each layer $\ell$, every mode is mixed on a $50{:}50$ beam splitter with vacuum in the ancillary port. The leaves $j$ at the output originate either from $A$ or $B$; we indicate the origin by colouring the leaves in yellow ($m(j)=A$) or green ($m(j)=B$). Leaf indices $j$ and layer indices $\ell$ follow the conventions defined in Eqs.~\eqref{Eq:app-definitions} and \eqref{Eq:app-quadrature-z}. (b) Legend showing the optical elements used in the beam–splitter tree.
    }
    \label{F-proof}
\end{figure*}

\section{Covariances and quantum Fisher information matrix of a two-mode squeezed state through a balanced tree \label{App:covariance-two-mode-squeezed-state-QFIM}}

In this appendix we derive closed-form expressions for the covariance matrix of a state that starts as a two-mode squeezed state and is then split into $M=2^N$ modes ($N\ge 2$) by a balanced beam splitter tree. At each layer, every mode is mixed on a $50{:}50$ beam splitter with vacuum in the ancillary port (see Fig.~\ref{F-proof}). We then give the covariance matrix after $2^N$ local phase shifts are encoded on the modes. Finally, we obtain an analytical form of the quantum Fisher information matrix, together with its eigenvalues and eigenvectors. The entire appendix can also be reproduced using a \texttt{Mathematica} notebook (see Ref.~\cite{adeoliveirajunior2025d}), where all matrices are provided and the results can be verified both analytically and numerically. For clarity, the discussion is organised into three parts.

\subsection{Compact expression for the beam–splitter tree covariance matrix} \label{App:covariance-matrix-tree}

In this subsection we derive compact formulas for the ``leaf–to–leaf'' covariances $\sigma^{(x)}_{jk}$ and $\sigma^{(p)}_{jk}$ generated by a balanced beam–splitter tree fed by a two-mode squeezed state on inputs $A,B$. The key idea is that each output quadrature $\hat z_j$ can be written as a scaled version of one of the original input quadratures ($A$ or $B$) together with extra terms coming from the vacuum modes that enter the unused ports of the beam splitters. These extra terms have simple $\pm$ signs that depend on the path through the tree. We will keep track of these signs and scaling factors, compute the covariances coming from the squeezed inputs and the vacua separately, then add them to obtain explicit formulas for $\sigma^{(x)}_{jk}$ and $\sigma^{(p)}_{jk}$. Finally, we combine them into the global block-diagonal covariance matrix $\v\sigma=\v\sigma^{(x)}\oplus\v\sigma^{(p)}$.

For a mode $m$ mixed with an ancillary vacuum mode $v$, a balanced beam splitter acts on the dimensionless quadratures
\begin{equation}\label{Eq:app-beam-splitter-action}
\begin{pmatrix} x_{m}\\ x_{v}\end{pmatrix}\longrightarrow
\frac{1}{\sqrt{2}}
\begin{pmatrix}1 & 1\\ 1 & -1\end{pmatrix}
\begin{pmatrix}x_{m}\\ x_{v}\end{pmatrix},
\qquad
\begin{pmatrix}p_{m}\\ p_{v}\end{pmatrix}\longrightarrow
\frac{1}{\sqrt{2}}
\begin{pmatrix}1 & 1\\ 1 & -1\end{pmatrix}
\begin{pmatrix}p_{m}\\ p_{v}\end{pmatrix}.
\end{equation}
Because \eqref{Eq:app-beam-splitter-action} is block-diagonal in $(x,p)$, it creates no cross $x$–$p$ correlations provided the input state itself is free of such terms. We adopt this assumption throughout.

Let $A$ and $B$ denote the two modes that carry an initial two‑mode squeezed state; every remaining input port is prepared in vacuum. A balanced binary tree of $50{:}50$ beam splitters of depth $N$ doubles the mode count at each layer. We count layers starting from 1; hence a tree of depth $N$ contains $N-1$ beam‑splitter layers and produces $M=2^{N}$ output modes (see Fig.~\ref{F-proof}).

We label leaves and layers by $j\in\{0,1,\dots,2^{N}-1\}$ and $\ell\in\{1,\dots,N-1\}$, and set
\begin{align}\label{Eq:app-definitions}
j_{i}(j) &:=\bigl\lfloor j/2^{\,i}\bigr\rfloor\bmod 2 \quad(i\in\mathbbm{N}_{0}),\\ 
r(j) &:=j_{N-1}(j),\\ 
s_{\ell}(j) &:=(-1)^{\,j_{N-\ell-1}(j)},\\ 
b(j,\ell) &:=\bigl\lfloor j/2^{\,N-\ell}\bigr\rfloor . \end{align}
The following lemma gives an explicit expression for the quadrature operator at each output mode in terms of the initial inputs and the vacuum contributions introduced at each layer of the beam splitter tree:

\begin{lem}\label{lem:leaf-quadrature}
For any $z\in\{x,p\}$ and every leaf $j$,
\begin{equation}\label{Eq:app-quadrature-z}
  \hat{z}_{j}
   = 2^{-(N-1)/2}\,\hat{z}_{m(j)}
   + \sum_{\ell=1}^{N-1}
       2^{-(N-\ell)/2}\,
       s_{\ell}(j)\,
       \hat{z}_{\mathrm{vac}}^{(\ell,b(j,\ell))},
\end{equation}
where $m(j)=A$ if $r(j)=0$ and $m(j)=B$ if $r(j)=1$.
\end{lem}
\begin{proof}
The proof proceeds by induction on the network depth $N$. For the base case $N=2$, the tree contains two beam splitters, each mixing one of $A,B$ with a fresh vacuum. Applying \eqref{Eq:app-beam-splitter-action} once gives
\begin{equation}
    \hat{z}_{j}=2^{-1/2}\,\hat{z}_{m(j)}+2^{-1/2}\,s_{1}(j)\,\hat{z}_{\mathrm{vac}}^{(1,b(j,1))},
\end{equation}
which matches \eqref{Eq:app-quadrature-z} for $N=2$. Assume \eqref{Eq:app-quadrature-z} holds for depth $N-1$. Attach to every leaf a balanced beam splitter that mixes the leaf with a fresh vacuum $\hat{z}_{\mathrm{vac}}^{(N-1,b)}$ ($b=0,\dots,2^{N-1}-1$). Denote by $\hat{z}^{\text{(old)}}_{j}$ the quadrature on the depth‑$(N-1)$ leaf $j$ before the extra splitter is inserted. The two new leaves, labelled $2j$ and $2j+1$, obey
\begin{equation}
    \hat{z}_{2j+\nu} =2^{-1/2}\Bigl[\hat{z}^{\text{(old)}}_{j}+(-1)^{\nu}\hat{z}_{\mathrm{vac}}^{(N-1,b)} \Bigr] \quad\text{for}\quad \nu\in\{0,1\}.
\end{equation}
Inserting the induction hypothesis for $\hat{z}^{\text{(old)}}_{j}$ and observing that $(-1)^{\nu}=s_{N-1}(2j+\nu)$ yields \eqref{Eq:app-quadrature-z} for depth $N$. The identities $b(2j+\nu,\ell)=b(j,\ell)$ and $s_\ell(2j+\nu)=s_\ell(j)$ for $\ell<N-1$ follow directly from the definitions, completing the proof.
\end{proof}
Next, we use the previous lemma to decompose each output quadrature into its squeezed and vacuum contributions, in preparation for computing covariances between the leaf modes. Define
\begin{equation}
\hat{z}_{j}^{(\mathrm{sq})}:=2^{-(N-1)/2}\,\hat{z}_{m(j)}\quad\text{and} \quad\hat{z}_{j}^{(\mathrm{vac})}:=\sum_{\ell=1}^{N-1}2^{-(N-\ell)/2}\,s_{\ell}(j)\,\hat{z}_{\mathrm{vac}}^{(\ell,b(j,\ell))}.
\end{equation}
Because the initial TMSS pair $(A,B)$ and all ancilla vacua are
uncorrelated and the network is a real orthogonal (hence
symplectic) transformation on $(x,p)$, cross terms remain zero:
$\langle\hat{z}_{j}^{(\mathrm{sq})}\hat{z}_{k}^{(\mathrm{vac})}\rangle =0$ for all $j,k$. With $\hat{x}=(\hat{a}+\hat{a}^\dagger)/\sqrt{2}$ and $\hat{p}=(\hat{a}-\hat{a}^\dagger)/(i\sqrt{2})$, the TMSS state satisfies $\operatorname{cov}(\hat{x}_A,\hat{x}_B)=-\tfrac12\sinh 2r$ and $\operatorname{cov}(\hat{p}_A,\hat{p}_B)=\tfrac12\sinh 2r$.  Propagating through the tree gives
\begin{align}
\operatorname{cov}\bigl(\hat{x}^{(\mathrm{sq})}_j,
                        \hat{x}^{(\mathrm{sq})}_k\bigr)
 &=\frac{1}{2^N}
   \bigl[\cosh 2r\,\delta_{m(j),m(k)}
         -\sinh 2r\,(1-\delta_{m(j),m(k)})\bigr],\label{Eq:sq-x}\\[6pt]
\operatorname{cov}\bigl(\hat{p}^{(\mathrm{sq})}_j,
                        \hat{p}^{(\mathrm{sq})}_k\bigr)
 &=\frac{1}{2^N}
   \bigl[\cosh 2r\,\delta_{m(j),m(k)}
         +\sinh 2r\,(1-\delta_{m(j),m(k)})\bigr].\label{Eq:sq-p}
\end{align}
For the vacuum ancillas, we have $\langle\hat{z}_{\mathrm{vac}}^{(\ell,b)} \hat{z}_{\mathrm{vac}}^{(\ell',b')}\rangle =\tfrac12\delta_{\ell,\ell'}\delta_{b,b'}$, so
\begin{equation}\label{Eq:vac-cov}
  \operatorname{cov}\bigl(\hat{z}^{(\mathrm{vac})}_j,
                          \hat{z}^{(\mathrm{vac})}_k\bigr)
  =\frac{1}{2}\sum_{\ell=1}^{N-1}
          2^{-(N-\ell)}\,
          s_\ell(j)\,s_\ell(k)\,
          \delta_{b(j,\ell),\,b(k,\ell)}.
\end{equation}
Adding~\eqref{Eq:sq-x}–\eqref{Eq:vac-cov} yields
\begin{align}
\sigma^{(x)}_{jk} &=
  2^{-N}\bigl[\cosh 2r\,\delta_{m(j),m(k)}
        -\sinh 2r\,(1-\delta_{m(j),m(k)})\bigr]
  +\frac{1}{2}\sum_{\ell=1}^{N-1}
          2^{-(N-\ell)}\,
          s_\ell(j)\,s_\ell(k)\,
          \delta_{b(j,\ell),\,b(k,\ell)},\label{Eq:sigma-x}\\[6pt]
\sigma^{(p)}_{jk} &=
  2^{-N}\bigl[\cosh 2r\,\delta_{m(j),m(k)}
        +\sinh 2r\,(1-\delta_{m(j),m(k)})\bigr]
  +\frac{1}{2}\sum_{\ell=1}^{N-1}
          2^{-(N-\ell)}\,
          s_\ell(j)\,s_\ell(k)\,
          \delta_{b(j,\ell),\,b(k,\ell)},\label{Eq:sigma-p}
\end{align}
which are symmetric and positive‑definite. Now we collect the $x$-and $p$-covariances into the full $2M\times 2M$ covariance matrix. Because the $\hat{x}$ and $\hat{p}$ quadratures remain uncorrelated throughout the network, the global covariance matrix takes the block-diagonal form
\begin{equation}\label{Eq:app-covariance-matrix-tree}
\v{\sigma} = \v{\sigma}^{(x)} \oplus \v{\sigma}^{(p)},
\end{equation}
with the canonical ordering $(\hat{x}_0, \hat{x}_1, \dots, \hat{x}_{M-1}, \hat{p}_0, \hat{p}_1, \dots, \hat{p}_{M-1})$. Alternatively, using the interleaved ordering $(\hat{x}_0, \hat{p}_0, \hat{x}_1, \hat{p}_1, \dots)$, we define a permutation $\mathcal{P}$ such that
\begin{equation}
\v{\sigma} =
\mathcal{P}\begin{pmatrix}
  \v{\sigma}^{(x)} & 0 \\
  0 & \v{\sigma}^{(p)}
\end{pmatrix}\mathcal{P}^\top,
\end{equation}
where $\mathcal{P}(2j)=j$, $\mathcal{P}(2j+1)=M+j$ for $j=0,\dots,M-1$. This completes the construction of the full covariance matrix of the output state. 

\subsection{Covariance matrix phase-shift}\label{App:covariance-matrix-phase-shift}

In this subsection we describe how the covariance matrix changes when local phase shifts are applied to each output mode of the beam–splitter tree. Each phase shift $\theta_j$ acts as a rotation in the $(x,p)$ quadrature space of mode $j$, and all such rotations together form a block–diagonal symplectic matrix $S(\v\theta)$. By applying the standard transformation rule $\v\sigma\mapsto S(\v\theta),\v\sigma,S(\v\theta)^\top$ to the block–diagonal covariance from the previous section, we obtain explicit formulas for the rotated covariance blocks $\v\sigma^{(xx)}(\v\theta)$, $\v\sigma^{(xp)}(\v\theta)$, $\v\sigma^{(px)}(\v\theta)$, and $\v\sigma^{(pp)}(\v\theta)$.

Consider the zero-mean Gaussian state produced by the beam splitter tree of depth $N$ given by Eq.~\eqref{Eq:app-covariance-matrix-tree}. Let $\v \theta =(\theta_0, ..., \theta_{M-1}) \in \mathbbm{R}^M$ and define the symplectic matrix:
\begin{equation}
    S(\v \theta) = \bigoplus_{j=0}^{M-1}\begin{pmatrix}
        \cos \theta_j & \sin \theta_j \\ -\sin \theta_j & \cos\theta_j
    \end{pmatrix}.
\end{equation}
The local phase-rotation unitary $\hat{U}(\v\theta) = \Pi_{j=0}^{M-1} e^{-i\theta_j \hat{a}^{\dagger}_j \hat a_j}$ transforms the covariance matrix as 
\begin{equation}\label{Eq:covariance-matrix-phase-shift}
    \v \sigma(\v\theta):= S(\v\theta) \v\sigma S(\v\theta)^\top = \begin{pmatrix}
        \v\sigma^{(xx)}(\v\theta) & \v\sigma^{(xp)}(\v\theta) \\ \v\sigma^{(px)}(\v \theta) & \v\sigma^{(pp)}(\v\theta)
    \end{pmatrix},
\end{equation}
with entries for $j,k=0,...,M-1$,
\begin{align}
    \sigma^{(xx)}_{jk}(\v\theta) &=\cos\theta_j\cos\theta_k\sigma_{jk}^{(x)}+\sin\theta_j\sin\theta_k\sigma_{jk}^{(p)}, \\
     \sigma^{(xp)}_{jk}(\v\theta) &= -\cos\theta_j\sin\theta_k\sigma_{jk}^{(x)}+\sin\theta_j\cos\theta_k\sigma_{jk}^{(p)},\\
      \sigma^{(px)}_{jk}(\v\theta) &=-\sin\theta_j\cos\theta_k\sigma_{jk}^{(x)}+\sin\theta_j\sin\theta_k\sigma_{jk}^{(p)}, \\
       \sigma^{(pp)}_{jk}(\v\theta) &=\sin\theta_j\sin\theta_k\sigma_{jk}^{(x)}+\cos\theta_j\cos\theta_k\sigma_{jk}^{(p)}. \\
\end{align}

The form of Eq.~\eqref{Eq:covariance-matrix-phase-shift} is easily understood once one recalls how a phase-rotation acts in the Heisenberg picture. First, for each mode $j$,
\begin{equation}
    \hat{U}(\v\theta)^{\dagger}\begin{pmatrix}
        \hat{x}_j \\ \hat{p}_j
    \end{pmatrix}U(\v\theta) = \begin{pmatrix}
        \cos \theta_j & \sin \theta_j \\ -\sin \theta_j & \cos\theta_j
    \end{pmatrix}\begin{pmatrix}
        \hat{x}_j \\ \hat{p}_j
        \end{pmatrix}.
\end{equation}
Because a linear (and, in particular, symplectic) transformation $S$ carries the covariance matrix to $\v \sigma \mapsto S(\v\theta)\v\sigma S(\v\theta)^\top$, the full transformation law follows at once.  Finally, substituting the block-diagonal form of the original $\v\sigma$ and the block structure of $S(\v\theta)$ yields the explicit block expression displayed in Eq.~\eqref{Eq:covariance-matrix-phase-shift}.

\subsection{Quantum Fisher information from a beam splitter tree} \label{App:QFIM-beam-splitter-tree}

We now compute the quantum Fisher information matrix for estimating the local phases $\v{\theta}$ applied to the modes at the output of the beam–splitter tree, and determine its eigenvalues and eigenvectors in closed form. Starting from the covariances $\sigma^{(x)}_{jk}$ and $\sigma^{(p)}_{jk}$ derived earlier, we first show that the QFIm does not depend on the actual phase values, so it can be evaluated at $\v{\theta}=\v{0}$. We then use the symmetry of the tree to prove that $\v Q$ has only three distinct kinds of entries, which allows us to write it in a simple projector form. This structure makes the eigendecomposition straightforward: two special phase patterns ($\v{u}_+$ for a uniform phase and $\v{u}_0$ for an $A$–vs–$B$ imbalance) give two of the eigenvalues, and any vector orthogonal to both lies in a degenerate subspace with a common eigenvalue $\lambda_\perp$. 

Consider the zero-mean Gaussian output of the depth-$N$
beam–splitter tree fed by a two-mode squeezed state on inputs $A,B$ (all other inputs in vacuum), with
quadrature covariances $\sigma^{(x)}_{jk},\sigma^{(p)}_{jk}$ given by Eqs.~\eqref{Eq:sigma-x}–\eqref{Eq:sigma-p}. Below, we prove that after encoding local phase shifts as in Eq.~\eqref{Eq:covariance-matrix-phase-shift}, the QFIm $\v Q$:
\begin{enumerate}[label=(\alph*)]
\item is independent of $\v\theta$ and has entries
\begin{equation}\label{Eq:Q-from-sigmas-final}
Q_{jk}=2\big[(\sigma^{(x)}_{jk})^2+(\sigma^{(p)}_{jk})^2\big]-\delta_{jk};
\end{equation}
\item admits the projector form
\begin{equation}\label{Eq:Q-final-proj}
\v Q=\lambda_\perp\bigl(\mathbbm{1}_M-\Pi_+-\Pi_0\bigr)+\lambda_+\,\Pi_+ \quad \text{with} \quad 
\Pi_+:=\frac{\v 1\v 1^\top}{M}\quad \text{and} \quad \Pi_0:=\frac{\v s\v s^\top}{M},
\end{equation}
where $\v 1=(1,\ldots,1)^\top$ and $\v s=(\underbrace{1,\ldots,1}_{\tfrac{M}{2}},\underbrace{-1,\ldots,-1}_{\tfrac{M}{2}})^\top$;
\item has spectrum
\begin{equation}\label{Eq:Q-spectrum-final}
\operatorname{\lambda(\v Q)} = \qty{0, \underbrace{\frac{\cosh 2r - 1}{2^{N-2}},...,\frac{\cosh 2r - 1}{2^{N-2}}}_{M-2 \: \text{times}},\frac{\cosh 4r -1}{2^{N-1}} }.
\end{equation}
\end{enumerate}
Equivalently, entrywise,
\begin{equation}\label{Eq:Q-casewise-final}
Q_{jk}=
\begin{cases}
\displaystyle \lambda_\perp+\frac{\lambda_+-2\lambda_\perp}{M}, & j=k,\\[6pt]
\displaystyle \frac{\lambda_+-2\lambda_\perp}{M}, & j \neq k \ \text{and}\ m(j)=m(k),\\[6pt]
\displaystyle \frac{\lambda_+}{M}, & m(j)\neq m(k),
\end{cases}
\end{equation}
where 
\begin{align}
    \lambda_\perp&=\frac{\cosh 2r-1}{2^{N-2}}\quad   \text{and} \quad \lambda_+=\frac{\cosh 4r-1}{2^{N-1}}.
\end{align}

\noindent
\emph{Step 1 (Independence of $\v\theta$ and entrywise formula).} By Lemma~\ref{Lem:QFIm} and zero mean, the QFIm is 
\begin{equation}\label{Eq:QFIM-zero-mean-again}
Q_{jk}(\v\theta)=2\sum_{z,w\in\{x,p\}}\!\sigma_{\hat z_j \hat w_k}(\v\theta)^2-\delta_{jk}, 
\end{equation} where $\sigma_{\hat z_j \hat w_k}(\v\theta)$ are the second moments after encoding. Using Frobenius‑norm invariance under left/right orthogonal multiplications~\cite{BengtssonZyczkowski2006},
\begin{align}
\sum_{z,w}\sigma_{\hat z_j \hat w_k}(\v\theta)^{2}
&=\bigl\|\v\sigma_{jk}(\v\theta)\bigr\|_F^2
= \bigl\|R(\theta_j)\,\v\sigma_{jk}(\v 0)\,R(\theta_k)^\top\bigr\|_F^2 = \bigl\|\v\sigma_{jk}(\v 0)\bigr\|_F^2
= (\sigma^{(x)}_{jk})^2+(\sigma^{(p)}_{jk})^2. \label{Eq:Frob-invariance}
\end{align}
Thus $Q_{jk}(\v\theta)$ is independent of $\v\theta$, and evaluating at $\v\theta=\v 0$ yields
\eqref{Eq:Q-from-sigmas-final}.

\emph{Step 2 (Projector form and commutant).} Let $v_{jk}$ denote the vacuum contribution appearing in \eqref{Eq:sigma-x}–\eqref{Eq:sigma-p}: 
\begin{equation}
    v_{jk}:=\frac{1}{2}\sum_{\ell=1}^{N-1}2^{-(N-\ell)}\,s_\ell(j)\,s_\ell(k)\,\delta_{b(j,\ell),\,b(k,\ell)}.
\end{equation}
We claim that
\begin{equation}\label{Eq:vac-three-values}
v_{jk}= \begin{cases}
 \frac{1}{2}\bigl(1-2^{\,1-N}\bigr), & j=k,\\[4pt]
-\,2^{-N}, & j \neq k \ \text{and}\ m(j)=m(k),\\[4pt]
 0, & m(j)\neq m(k).
\end{cases}
\end{equation}
For $j=k$, the signs coincide $s_\ell(j)=s_\ell(k)$ and all Kronecker deltas are $1$, so $v_{jj}=\tfrac12\sum_{\ell=1}^{N-1}2^{-(N-\ell)} =\tfrac12(1-2^{1-N})$. If $m(j)\neq m(k)$, then $j$ and $k$ lie on opposite outputs of the first (top) beam splitter, so $b(j,1)\neq b(k,1)$. At every subsequent level $\ell\ge 2$, each side is split further without mixing with the other, hence $b(j,\ell)\neq b(k,\ell)$ for all $\ell$. Therefore $\delta_{b(j,\ell),b(k,\ell)}=0$ for every $\ell$, and $v_{jk}=0$. Now take $j\neq k$ with $m(j)=m(k)$. Let $L\in\{2,\ldots,N-1\}$ be the smallest level at which $j$ and $k$ choose different outputs (equivalently, the last level where they coincide is $L-1$). Then only the terms with $\ell=1,\ldots,L-1$ contribute to $v_{jk}$. For $\ell\le L-2$ their choices agree, so $s_\ell(j)s_\ell(k)=+1$, whereas at $\ell=L-1$ they differ and $s_{L-1}(j)s_{L-1}(k)=-1$. Hence,
\begin{equation}
    v_{jk}=\frac12\qty[\sum_{\ell=1}^{L-2}2^{-(N-\ell)}-2^{-(N-L+1)}] =\frac12\qty[2^{-(N-L+1)}(1-2^{-(L-2)})-2^{-(N-L+1)}]
=-\,2^{-N},
\end{equation}
proving Eq.~\eqref{Eq:vac-three-values}. 

According to Eq.~\eqref{Eq:vac-three-values} and the structure of the squeezed part in \eqref{Eq:sigma-x}–\eqref{Eq:sigma-p}, the entries of $Q_{jk}$ in \eqref{Eq:Q-from-sigmas-final} take only three values: one for $j=k$, one for $j\neq k$ with $m(j)=m(k)$, and one for $m(j)\neq m(k)$. Hence $\v Q$ is completely determined by three numbers and can be written as
\begin{equation}\label{Eq:Q-projector-final}
\v Q=\alpha\,\mathbbm{1}_M+\beta\,\Pi_+ + \gamma\,\Pi_0,
\end{equation}
for some scalars $\alpha,\beta,\gamma$, where $\Pi_+=\v 1\v 1^\top/M$ and $\Pi_0=\v s\v s^\top/M$, with
\begin{equation}
    \v 1\v 1^\top=\begin{pmatrix}\v 1_A\v 1_A^\top&\v 1_A\v 1_B^\top\\ \v 1_B\v 1_A^\top&\v 1_B\v 1_B^\top\end{pmatrix} \quad \text{and} \quad \v s\v s^\top=\begin{pmatrix}\v 1_A\v 1_A^\top&-\v 1_A\v 1_B^\top\\ -\v 1_B\v 1_A^\top&\v 1_B\v 1_B^\top\end{pmatrix}.
\end{equation}
where $\v 1_A$ and $\v 1_B$ denote the all-ones vectors of length $\tfrac{M}{2}$ on the $A$ and $B$ indices, respectively.

\emph{Step 3 (Eigenspaces and eigenvalues).}
We now pick two phase patterns that match the symmetry of the tree and make the QFIm simple to evaluate. If we vary the phases along a unit vector $\v u\in\mathbb R^M$ via $\v\theta=t\,\v u$, the unitary is
$\hat U(t)=\exp\!\big(-it\,\hat G_{\!u}\big)$ with generator
\begin{equation}
    \hat G_{\!u}:=\sum_{j=0}^{M-1} u_j\,\hat n_j.
\end{equation}
For our zero-mean Gaussian setting, one has $\v u^\top \v Q\,\v u \;=\; 4\,\mathrm{var}(\hat G_{\!u})$, so we want directions $\v u$ for which $\hat G_{\!u}$ has a simple form on our state. Define,
\begin{equation}
    \v u_+=\frac{\v 1}{\sqrt{M}}
\quad\text{and}\quad
\v u_0=\frac{\v s}{\sqrt{M}},
\qquad
\text{where } s_j=\begin{cases}+1,& j\in A\\ -1,& j\in B.\end{cases}
\end{equation}
These correspond to a uniform phase (all modes together) and an $A$-vs-$B$ imbalance (plus on $A$, minus on $B$). A passive beam–splitter network redistributes photons but does not create or destroy them. Since all ancillary inputs (other than $A,B$) are in vacuum, the total photon number flowing down the $A$-branch equals the photon number that entered at $A$, and similarly for $B$:
\begin{equation}
    \sum_{j\in A}\hat n_j=\hat n_A,
    \qquad
    \sum_{k\in B}\hat n_k=\hat n_B,
    \qquad
    \sum_{j=0}^{M-1}\hat n_j=\hat n_A+\hat n_B.
\end{equation}
Consequently, the generators for $\v u_+$ and $\v u_0$ collapse to
\begin{equation}
    \hat G_+=\sum_j \frac{1}{\sqrt{M}}\hat n_j
    =\frac{\hat n_A+\hat n_B}{\sqrt{M}}\:\:,\:\:
    \hat G_0=\sum_j \frac{s_j}{\sqrt{M}}\hat n_j
    =\frac{\hat n_A-\hat n_B}{\sqrt{M}}.
\end{equation}
These simple number combinations make the corresponding variances, and hence the eigenvalues of $\v Q$ along $\v u_+$ and $\v u_0$, straightforward to compute for a TMSS. For a TMSS with squeezing $r$ and $\bar n:=\sinh^2 r$, $\operatorname{var}(\hat n_A)=\operatorname{var}(\hat n_B)=\bar n(\bar n+1),\:
\operatorname{cov}(\hat n_A,\hat n_B)=\bar n(\bar n+1)$, so $\operatorname{var}(\hat n_A+\hat n_B)=4\,\bar n(\bar n+1)$ and $\operatorname{var}(\hat n_A-\hat n_B)=0$. For our pure state under local phase shifts, the QFIm satisfies along any direction $\v u$: $\v u^{\top} \v Q \v u = 4\operatorname{var}(\hat{G}_u)$. Since $\v u_+$ and $\v u_0$ are normalised and are eigenvectors by symmetry, their eigenvalues are $\lambda_\pm = 4\operatorname{var}(\hat{G}_\pm)$. Computing then, we have that for $\v u_+$:
\begin{equation}
    \operatorname{var}(\hat{G}_+) = \frac{1}{M}\operatorname{var}(\hat{n}_A+\hat{n}_B) = \frac{4\bar{n}(\bar{n}+1)}{M} = \frac{\cosh 4r-1}{2^{N-1}}.
\end{equation}
For $\v u_0$, it follows that
\begin{equation}
    \operatorname{var}(\hat{G}_0) = \frac{1}{M}\operatorname{var}(\hat{n}_A-\hat{n}_B) = 0.
 \end{equation}

Finally, from the three-valued structure of $\v Q_{jk}$, the QFIm lies in the three-dimensional linear span generated by $\{\v 1, \Pi_+, \Pi_0 \}$. Consequently, $\mathbbm{R}^M$ decomposes into three invariant (mutually orthogonal) subspaces: $\operatorname{span}\{\v u_+\}\oplus\operatorname{span}\{\v u_0\}\oplus \mathcal{W}$, where $\mathcal{W}:=\{\v w:\ \v 1^\top \v w=0,\ \v s^\top \v w=0\}$. On $\mathcal{W}$, we have  $\v \Pi_+ \v w = \v \Pi_0 \v w = 0$. So, $\v Q \v w = \lambda_{\perp}\v w$. To obtain the eigenvalue, choose $\v w$ supported on two sibling leaves $j,k$ in the same half created by the last balanced splitter, with entries $w_j=1$, $w_k=-1$, others $0$. The associated generator is $\hat G_\perp=\frac{\hat n_j-\hat n_k}{\sqrt{2}}$. Writing the last splitter as $\hat a_j=\tfrac{1}{\sqrt{2}}(\hat c+\hat v)$, $\hat a_k=\tfrac{1}{\sqrt{2}}(\hat c-\hat v)$ with $\hat v$ a fresh vacuum and $\hat c$ the parent mode one layer above, one finds $\hat n_j-\hat n_k=\hat c^\dagger \hat v+\hat v^\dagger \hat c$. On a product state $\rho_c\otimes\ket0\!\bra0_v$ (i.e.\ $\hat v$ in vacuum and uncorrelated with $\hat c$),
\begin{equation}\label{Eq:var-diff}
\mathrm{var}(\hat n_j-\hat n_k)
=\big\langle (\hat c^\dagger \hat v+\hat v^\dagger \hat c)^2\big\rangle
=\langle \hat c^\dagger \hat c\rangle,
\end{equation}
since $\langle \hat v^\dagger \hat v\rangle=0$, $\langle \hat v \hat v^\dagger\rangle=1$ and all mixed
expectations factorise.
Therefore
\begin{equation}\label{Eq:lambda-perp-final}
\lambda_\perp
=\frac{\v w^\top \v Q\,\v w}{\|\v w\|^2}
=4\,\mathrm{var}(\hat G_\perp)
=2\,\mathrm{var}(\hat n_j-\hat n_k)
\overset{\eqref{Eq:var-diff}}=2\,\langle \hat c^\dagger \hat c\rangle.
\end{equation}
Each $50{:}50$ splitter with vacuum halves the mean photon number on the transmitted branch: if $\hat a_{\rm out}=(\hat a_{\rm in}+\hat v)/\sqrt2$ with $\hat v$ in vacuum then $\langle \hat a_{\rm out}^\dagger \hat a_{\rm out}\rangle=\tfrac12\langle \hat a_{\rm in}^\dagger \hat a_{\rm in}\rangle$. Along any path from an input to a leaf there are $N-1$ splitters; consequently the leaf mean is $\bar n/2^{\,N-1}$. The parent mode $\hat c$ sits one layer above the leaves, so it has passed through $N-2$ splitters and $\langle \hat c^\dagger \hat c\rangle=\frac{\bar n}{2^{\,N-2}}$. Substituting into \eqref{Eq:lambda-perp-final} and recalling $\cosh 2r-1=2\bar n$ gives
\begin{equation}
    \lambda_\perp=\frac{2\bar n}{2^{\,N-2}}=\frac{\cosh 2r-1}{2^{\,N-2}}
\end{equation}
Therefore $\v Q$ has eigenvalue $0$ on $\mathrm{span}\{\v u_0\}$, eigenvalue $\lambda_+$ on $\mathrm{span}\{\v u_+\}$, and eigenvalue $\lambda_\perp$ on $\mathcal W$, which is equivalent to the projector form \eqref{Eq:Q-final-proj} and to the spectrum \eqref{Eq:Q-spectrum-final}. The casewise expression \eqref{Eq:Q-casewise-final} follows from \eqref{Eq:Q-final-proj} by evaluating the three cases $j=k$, $j \neq k$ with $m(j)=m(k)$, and $m(j)\neq m(k)$.

\subsection{Average number of photons hitting the phase-shift}\label{App:mean-photon-number}

In the previous appendix, we proved that the eigenvalues of the QFIm associated with the $2^N$-mode Gaussian state generated by recursively distributing a two-mode squeezed vacuum through a balanced beam splitter network exhibit a highly structured eigenspectrum. More precisely, these are given by
\begin{equation}\label{Eq:eigenvalues-F-2} 
    \operatorname{\lambda(\v Q)} = \qty{0, \underbrace{\frac{\cosh 2r - 1}{2^{N-2}},...,\frac{\cosh 2r - 1}{2^{N-2}}}_{2^N-2 \: \text{times}},\frac{\cosh 4r -1}{2^{N-1}} }.
\end{equation}
We now recast them in terms of the mean photon number. The mean photon number per input mode is $\bar n:= \sinh^2 r$, so the total mean photon number on the two-mode squeezed state is $\bar{N}:=2\sinh^2 r$. After the balanced $2^N$-port splitter network, this total is conserved and spread uniformly, hence the mean photons per output mode (i.e., per local phase parameter) is
\begin{equation}\label{Eq:average-photon-number-n}
    \bar{n}_{\text{mode}} = \frac{\bar{N}}{2^N} = 2^{1-N} \sinh^2 r.
\end{equation}
Using the identities $\cosh 2r -1 = 2\sinh^2r$ and $\cosh 4r -1 = 2\sinh^2 2r = 8\sinh^2 r \cosh ^2 r$, the spectrum can be written as
\begin{equation}
    \lambda(\v Q) = \qty{0, \underbrace{4\bar{n}_{\text{mode}}, ...,4\bar{n}_{\text{mode}}}_{2^{N}-2 \:\text{times}}, 8\bar{n}_{\text{mode}}(1+2^{N-1}\bar{n}_{\text{mode}})}.
\end{equation}
For fixed $N$, the QFIm spectrum splits into three sectors: one null direction, a $(2^N-2)$-dimensional shot-noise sector with eigenvalues $\lambda_{\text{SNL}}:=4\bar n_{\text{mode}}$ that grow linearly with the mean photons per mode, and one enhanced direction with $\lambda_{\max}:=8\bar n_{\text{mode}}\bigl(1+2^{N-1}\bar n_{\text{mode}}\bigr)$ that includes a quadratic term from squeezing-induced correlations. In terms of total photons $\bar N=2^N\bar n_{\text{mode}}$, these become $\lambda_{\text{SNL}}=\bar N/2^{N-2}$ and $\lambda_{\max}=\bar N^{2}/2^{N-2}+8\bar N/2^{N}$, so the enhancement factor is $\lambda_{\max}/\lambda_{\text{SNL}}=\bar N+2$. Consequently, estimation variances scale as $1/\lambda$: generic phase combinations are shot-noise limited $\sim 1/\bar N$, while the special collective phase achieves Heisenberg scaling $\sim 1/\bar N^{2}$. The factor $2^N$ only fixes prefactors for a given $N$--it does not change the photon-number scaling exponent. 

\section{Quantum Fisher information matrix \& privacy}\label{App:analytics-QFIm}

In this appendix, we prove Lemma~\ref{Lem:QFIm} and Theorem~\ref{Thm:privacy-n}. We begin by considering a pure state $\ket{\psi_0}$ and recalling that a local phase shift on mode $\mu$ is implemented by the unitary $U(\theta_\mu) = e^{i \theta_{\mu} \hat{n}_\mu}$, where $\hat{n}_\mu := \hat{a}^{\dagger}_\mu \hat{a}_\mu$ is the number operator. For pure states, the quantum Fisher information matrix (QFIm) is given by
\begin{equation}\label{Eq:app-Quantum-Fisher-information-pure}
    \v{Q}_{\mu \nu}(\v \theta) = 4\text{Re}\qty[\braket{\partial_{\theta_{\mu}} \Psi(\v \theta)}{\partial_{\theta_{\nu}} \Psi(\v \theta)}-\braket{\partial_{\theta_\mu} \Psi(\v \theta)}{\Psi(\v \theta)}\braket{\Psi(\v \theta)}{\partial_{\theta_\nu} \Psi(\v \theta)}]
\end{equation}
Since the local phase-shift unitaries acting on different modes commute, the derivative with respect to each  $\theta_{\mu}$ simplifies to:
\begin{equation}\label{Eq:app-nice-relation}
    \ket{ \partial_{\theta_\mu}\Psi(\v \theta)} = \frac{\partial}{\partial \theta_{\mu}}\qty(e^{i\theta_\mu \hat{a}^{\dagger}_{\mu} \hat{a}_{\mu}})\qty(\prod_{\nu \neq \mu} e^{i\theta \hat{a}^{\dagger}_{\nu} \hat{a}_{\nu}}) \ket{\psi_0} = i \hat{n}_{\mu} \ket{\Psi(\v \theta)}. 
\end{equation}
Plugging Eq.~\eqref{Eq:app-nice-relation} into Eq.~\eqref{Eq:app-Quantum-Fisher-information-pure}, we find that the QFIm is given by the covariance of the number operators $\mu$ and $\nu$, i.e.,
\begin{equation}\label{Eq:app-QFI-4cov}
    \v{Q}_{\mu \nu}(\v \theta) = 4\qty[\langle \hat{n}_{\mu} \hat{n}_{\nu}\rangle-\langle \hat{n}_{\mu}\rangle\langle \hat{n}_{\nu}\rangle] = 4\,\text{cov}(\hat{n}_\mu, \hat{n}_{\nu})
\end{equation}
where all expectation values $\langle \bullet\rangle = \langle \Psi(\v \theta)|\bullet|\Psi(\v \theta)\rangle$ are taken with respect to the final state.

We now show that the covariance of number operators can be computed directly from the covariance matrix of the state. First, we express the number operator in terms of quadratures: $\hat{n}_{\mu} =\hat{a}^{\dagger}_\mu \hat{a}_\mu = \frac{1}{2}(\hat{x}_\mu -i \hat{p}_\mu)(\hat{x}_\mu +i \hat{p}_\mu)=\frac{1}{2}(\hat{x}_\mu^2 + \hat{p}_\mu^2-1)$, where $\hat{a} = (\hat{x}_\mu+i\hat{p}_\mu)/\sqrt{2}$. For zero-mean Gaussian states, all odd moments vanish, and fourth-order moments can be reduced using Wick's theorem~\cite{peskin2018introduction}
\begin{align}\label{Eq:app-wicks-theorem}
    \langle \hat{z}_{\alpha}\hat{z}_\beta \hat{z}_\gamma \hat{z}_\delta\rangle &=\langle \hat{z}_{\alpha}\hat{z}_\beta\rangle \langle \hat{z}_\gamma \hat{z}_\delta\rangle+\langle \hat{z}_{\alpha} \hat{z}_\gamma\rangle \langle \hat{z}_\beta \hat{z}_\delta\rangle + \langle \hat{z}_{\alpha} \hat{z}_\delta\rangle \langle \hat{z}_\beta \hat{z}_\gamma\rangle ,\nonumber \\
     \langle \hat{z}_{\alpha} \hat{z}_\beta \hat{w}_\gamma \hat{w}_\delta\rangle &=\langle \hat{z}_{\alpha} \hat{z}_\beta\rangle \langle \hat{w}_\gamma \hat{w}_\delta\rangle+\langle \hat{z}_{\alpha} \hat{w}_\gamma\rangle \langle \hat{z}_\beta \hat{w}_\delta\rangle + \langle \hat{z}_{\alpha} \hat{w}_\delta\rangle \langle \hat{z}_\beta \hat{w}_\gamma\rangle,
\end{align}
where $\hat z, \hat w \in \{\hat x,\hat p\}$. In particular, setting $\alpha = \beta= \mu$ and $\gamma = \delta  = \nu$, we obtain:
\begin{align}
    \langle \hat{z}^2_\mu \hat{z}^2_\nu \rangle &= \langle \hat{z}^2_\mu\rangle \langle \hat{z}^2_\nu \rangle+2\langle z_\mu z_\nu\rangle^2, \\ 
    \langle \hat{z}^2_\mu \hat{w}^2_\nu\rangle &= \langle \hat{z}^2_\mu\rangle \langle w^2_\nu \rangle+2\langle z_\mu w_\nu\rangle^2\label{Eq:wick-relation-2}.
\end{align}
Expanding the terms in $4\,\text{cov}(\hat{n}_\mu, \hat{n}_\nu)$ in terms of quadratures $\hat{x}_\mu, \hat{p}_\mu$, we find:
\begin{align}
     4\langle \hat{n}_\mu \hat{n}_\nu\rangle &= \langle(\hat{x}_\mu^2 + \hat{p}_\mu^2-1)(\hat{x}_\nu^2 + \hat{p}_\nu^2-1)\rangle= 1+ \langle \hat{x}_\mu^2 \hat{x}_\nu^2\rangle+\langle \hat{p}_\mu^2 \hat{p}_\nu^2\rangle+\langle \hat{x}_\mu^2 \hat{p}_\nu^2\rangle+\langle \hat{p}_\nu^2 \hat{x}_\mu^2\rangle - 2\langle \hat{p}_\mu^2\rangle-2\langle \hat{x}_\mu^2\rangle.\label{Eq:app-part-cov-1}\\
    4\langle \hat{n}_{\mu}\rangle\langle \hat{n}_{\nu}\rangle&=(\langle x^2_\mu+p^2_\mu-1\rangle\langle x^2_\nu+p^2_\nu-1\rangle) = 1+\langle x^2_\mu\rangle\langle x^2_\nu\rangle+\langle p^2_\mu\rangle\langle p^2_\nu\rangle+2\langle x^2_\mu\rangle\langle p^2_\nu\rangle -2\langle x^2_\mu\rangle -2\langle p^2_\mu\rangle\label{Eq:app-part-cov-2}.
\end{align}
For displaced Gaussian states $\hat{z}_{\alpha} \to \delta z_\alpha + \langle \hat{z}_{\alpha} \rangle$ with $\hat z\in \{\hat x,\hat p\}$ and $\alpha \in \{\mu, \nu\}$, we compute the following product:
\begin{align}
\hat{z}_\mu^2 \hat{w}_\nu^2 
&= \delta \hat{z}_\mu^2 \delta \hat{w}_\nu^2
+ 2 \langle \hat{w}_\nu \rangle\, \delta \hat{z}_\mu^2 \delta \hat{w}_\nu
+ \langle \hat{w}_\nu \rangle^2\, \delta \hat{z}_\mu^2 \nonumber \\
&\quad + 2 \langle \hat{z}_\mu \rangle\, \delta \hat{z}_\mu \delta \hat{w}_\nu^2
+ 4 \langle \hat{z}_\mu \rangle \langle \hat{w}_\nu \rangle\, \delta \hat{z}_\mu \delta \hat{w}_\nu
+ 2 \langle \hat{z}_\mu \rangle \langle \hat{w}_\nu \rangle^2\, \delta \hat{z}_\mu \nonumber \\
&\quad + \langle \hat{z}_\mu \rangle^2\, \delta \hat{w}_\nu^2
+ 2 \langle \hat{z}_\mu \rangle^2 \langle \hat{w}_\nu \rangle\, \delta \hat{w}_\nu
+ \langle \hat{z}_\mu \rangle^2 \langle \hat{w}_\nu \rangle^2. \label{eq:expanded-full}
\end{align}
Taking the expectation value and applying Wick’s theorem, we see that all odd fluctuation moments vanish, and the fourth-order moments decompose as:
\begin{align}
\langle\hat{z}_\mu^2 \hat{w}_\nu^2 \rangle= \langle\delta \hat{z}_\mu^2 \rangle \langle\delta \hat{w}_\nu^2\rangle
+2\langle \delta \hat{z}_\mu \delta \hat{w}_\nu\rangle^2 +\langle \hat{w}_\nu \rangle^2\, \langle\delta \hat{z}_\mu^2\rangle 
+ \langle \hat{z}_\mu \rangle^2\, \langle\delta \hat{w}_\nu^2\rangle
+4 \langle \hat{z}_\mu \rangle \langle \hat{w}_\nu \rangle\, \langle\delta \hat{z}_\mu \delta \hat{w}_\nu
\rangle+ \langle \hat{z}_\mu \rangle^2 \langle \hat{w}_\nu \rangle^2. \label{Eq:expanded-full}
\end{align}

We can now use Eq.~\eqref{Eq:expanded-full} to compute both terms of $4\operatorname{cov}(n_\mu,n_\nu)$. Starting with the first, we find that $4\langle \hat{n}_\mu \hat{n}_\nu\rangle$ is given by:
\begin{align}
4\langle \hat{n}_\mu \hat{n}_\nu\rangle 
= 1 
&+ \langle \delta \hat{x}_\mu^2 \rangle \langle \delta \hat{x}_\nu^2 \rangle
+ 2\langle \delta \hat{x}_\mu \delta \hat{x}_\nu \rangle^2
+ \langle \hat{x}_\nu \rangle^2 \langle \delta \hat{x}_\mu^2 \rangle
+ \langle \hat{x}_\mu \rangle^2 \langle \delta \hat{x}_\nu^2 \rangle
+ \langle \hat{x}_\mu \rangle^2 \langle \hat{x}_\nu \rangle^2 +4 \langle \hat{x}_\mu \rangle \langle \hat{x}_\nu \rangle\, \langle\delta \hat{x}_\mu \delta \hat{x}_\nu
\rangle\nonumber \\
 &+ \langle \delta \hat{p}_\mu^2 \rangle \langle \delta \hat{p}_\nu^2 \rangle
+ 2\langle \delta \hat{p}_\mu \delta \hat{p}_\nu \rangle^2
+ \langle \hat{p}_\nu \rangle^2 \langle \delta \hat{p}_\mu^2 \rangle
+ \langle \hat{p}_\mu \rangle^2 \langle \delta \hat{p}_\nu^2 \rangle
+ \langle \hat{p}_\mu \rangle^2 \langle \hat{p}_\nu \rangle^2+4 \langle \hat{p}_\mu \rangle \langle \hat{p}_\nu \rangle\, \langle\delta \hat{p}_\mu \delta \hat{p}_\nu
\rangle \nonumber \\
&+\langle \delta \hat{x}_\mu^2 \rangle \langle \delta \hat{p}_\nu^2 \rangle
+ 2\langle \delta \hat{x}_\mu \delta \hat{p}_\nu \rangle^2
+ \langle \hat{p}_\nu \rangle^2 \langle \delta \hat{x}_\mu^2 \rangle
+ \langle \hat{x}_\mu \rangle^2 \langle \delta \hat{p}_\nu^2 \rangle
+ \langle \hat{x}_\mu \rangle^2 \langle \hat{p}_\nu \rangle^2+4 \langle \hat{x}_\mu \rangle \langle \hat{p}_\nu \rangle\, \langle\delta \hat{x}_\mu \delta \hat{p}_\nu
\rangle\nonumber\\
&+\langle \delta \hat{p}_\mu^2 \rangle \langle \delta \hat{x}_\nu^2 \rangle
+ 2\langle \delta \hat{p}_\mu \delta \hat{x}_\nu \rangle^2
+ \langle \hat{x}_\nu \rangle^2 \langle \delta \hat{p}_\mu^2 \rangle
+ \langle \hat{p}_\mu \rangle^2 \langle \delta \hat{x}_\nu^2 \rangle
+ \langle \hat{p}_\mu \rangle^2 \langle \hat{x}_\nu \rangle^2+4 \langle \hat{p}_\mu \rangle \langle \hat{x}_\nu \rangle\, \langle\delta \hat{p}_\mu \delta \hat{x}_\nu
\rangle\nonumber
\\ &-2\langle \hat{x}^2_\mu \rangle-2\langle \hat{p}^2_\mu \rangle.
\end{align}
The second term, $4\langle n_\mu \rangle\langle n_\nu \rangle$, is given by:
\begin{align}
4\langle \hat{n}_\mu \rangle \langle \hat{n}_\nu \rangle =
1&+\langle \delta \hat{x}_\mu^2 \rangle \langle \delta \hat{x}_\nu^2 \rangle
+ \langle \delta \hat{x}_\mu^2 \rangle \langle \hat{x}_\nu \rangle^2
+ \langle \hat{x}_\mu \rangle^2 \langle \delta \hat{x}_\nu^2 \rangle
+ \langle \hat{x}_\mu \rangle^2 \langle \hat{x}_\nu \rangle^2  \nonumber\\
&+ \langle \delta \hat{p}_\mu^2 \rangle \langle \delta \hat{p}_\nu^2 \rangle
+ \langle \delta \hat{p}_\mu^2 \rangle \langle \hat{p}_\nu \rangle^2
+ \langle \hat{p}_\mu \rangle^2 \langle \delta \hat{p}_\nu^2 \rangle
+ \langle \hat{p}_\mu \rangle^2 \langle \hat{p}_\nu \rangle^2 \nonumber\\
&+ \langle \delta \hat{x}_\mu^2 \rangle \langle \delta \hat{p}_\nu^2 \rangle
+ \langle \delta \hat{x}_\mu^2 \rangle \langle \hat{p}_\nu \rangle^2
+ \langle \hat{x}_\mu \rangle^2 \langle \delta \hat{p}_\nu^2 \rangle
+ \langle \hat{x}_\mu \rangle^2 \langle \hat{p}_\nu \rangle^2 \nonumber\\
&+ \langle \delta \hat{p}_\mu^2 \rangle \langle \delta \hat{x}_\nu^2 \rangle
+ \langle \delta \hat{p}_\mu^2 \rangle \langle \hat{x}_\nu \rangle^2
+ \langle \hat{p}_\mu \rangle^2 \langle \delta \hat{x}_\nu^2 \rangle
+ \langle \hat{p}_\mu \rangle^2 \langle \hat{x}_\nu \rangle^2 \nonumber\\
&- 2\langle \delta \hat{x}_\mu^2 \rangle - 2\langle \hat{x}_\mu \rangle^2
- 2\langle \delta \hat{p}_\mu^2 \rangle - 2\langle \hat{p}_\mu \rangle^2.
\end{align}
Finally, by combining both terms, the covariance of the number operators can be expressed as:
\begin{align}
    4\operatorname{cov}(\hat{n}_\mu,\hat{n}_\nu) &= 2[\langle \delta \hat{x}_\mu \delta \hat{x}_\nu \rangle^2+\langle \delta \hat{p}_\mu \delta \hat{x}_\nu \rangle^2+\langle \delta \hat{x}_\mu \delta \hat{p}_\nu \rangle^2+\langle \delta \hat{p}_\mu \delta \hat{p}_\nu \rangle^2] \nonumber\\ &\hspace{0.5cm}+ 4\qty[\langle \hat{x}_\mu \rangle \langle \hat{x}_\nu \rangle\, \langle\delta \hat{x}_\mu \delta \hat{x}_\nu
\rangle+\langle \hat{p}_\mu \rangle \langle \hat{p}_\nu \rangle\, \langle\delta \hat{p}_\mu \delta \hat{p}_\nu
\rangle+\langle \hat{x}_\mu \rangle \langle \hat{p}_\nu \rangle\, \langle\delta \hat{x}_\mu \delta \hat{p}_\nu
\rangle+\langle \hat{p}_\mu \rangle \langle \hat{x}_\nu \rangle\, \langle\delta \hat{p}_\mu \delta \hat{x}_\nu
\rangle] - \delta_{\mu\nu},
\end{align}
 where the Kronecker delta $\delta_{\mu\nu}$ accounts for the non commutativity of the quadrature operators.
 Therefore, the quantum Fisher information can be simply written in terms of the covariance matrix and first moments as:
\begin{align}
\v Q_{\mu\nu} &= 2 \left(\sigma_{\hat{x}_\mu \hat{x}_\nu}^2 + \sigma_{\hat{p}_\mu \hat{p}_\nu}^2 + \sigma_{\hat{x}_\mu \hat{p}_\nu}^2 + \sigma_{\hat{p}_\mu \hat{x}_\nu}^2 \right) + 4 \Bigl(\sigma_{\hat{x}_\mu \hat{x}_\nu} \bar{r}_{\hat{x}_\mu} \bar{r}_{\hat{x}_\nu} + \sigma_{\hat{p}_\mu \hat{p}_\nu} \bar{r}_{\hat{p}_\mu}  \bar{r}_{\hat{p}_\nu}  + \sigma_{\hat{x}_\mu \hat{p}_\nu} \bar{r}_{\hat{x}_\mu}  \bar{r}_{\hat{p}_\nu}  + \sigma_{\hat{p}_\mu \hat{x}_\nu} \bar{r}_{\hat{p}_\mu}  \bar{r}_{\hat{x}_\nu}  \Bigr) - \delta_{\mu\nu} \nonumber\\ &= 2 \sum_{z,w \in \{x, p\}} \left( \sigma_{\hat{z}_\mu \hat{w}_\nu}^2 + 2\, \sigma_{\hat{z}_\mu \hat{w}_\nu} \, \bar{r}_{\hat{z}_\mu} \bar{r}_{\hat{w}_\nu} \right) - \delta_{\mu\nu},
\end{align}
where $\sigma_{\hat{z}_\alpha \hat{w}_\beta}:=\langle(\hat z_\alpha-\bar r_{\hat z_\alpha})(\hat w_\beta-\bar r_{\hat w_\beta})\rangle$ denotes the centred second moment and $\bar r_{\hat{z}_\alpha} = \langle \hat{z}_\alpha\rangle$. This concludes the proof of Lemma~\ref{Lem:QFIm}.

To prove Theorem~\ref{Thm:privacy-n}, we begin by recalling the definition of the privacy measure: $P(\v v_{\text{avg}}, \v Q) = \v v_{\text{avg}}^{\top} \v Q \v v_{\text{avg}} / \tr(\v Q)$, where $\v v_{\text{avg}} = (1, \dots, 1)^{\top} / \sqrt{M}$. Given the quantum Fisher information matrix $\v Q$ derived in Lemma~\ref{Lem:QFIm}, the numerator of $P$ can be written as

\begin{equation}
    \v v_{\text{avg}}^{\top}\v Q \v v_{\text{avg}} = \frac{1}{M}\sum_{\mu,\nu} Q_{\mu\nu} = \frac{2}{M}\sum_{\mu,\nu}\sum_{z,w\in\{x,p\}} \left( \sigma_{\hat z_\mu \hat w_\nu}^2 +2\sigma_{\hat z_\mu \hat w_\nu} \bar r_{\hat z_\mu}\bar r_{\hat w_\nu} \right)-1 .
\end{equation}
Similarly, the trace is given by $\tr(\v Q) = \sum_\mu Q_{\mu\mu}=2\sum_\mu \sum_{z,w \in \{x, p\}} \left( \sigma_{\hat{z}_\mu \hat{w}_\mu}^2 + 2\, \sigma_{\hat{z}_\mu \hat{w}_\mu} \, \bar{r}_{\hat{z}_\mu} \bar{r}_{\hat{w}_\mu} \right) - M$. Combining the numerator and the trace, we obtain the following closed-form expression for the privacy:
\begin{align}
    P(\v{Q}, \v{v}) =\frac{2\sum_{\mu,\nu}  \sum_{z,w \in \{x, p\}} \left( \sigma_{\hat{z}_\mu \hat{w}_\nu}^2 + 2\, \sigma_{\hat{z}_\mu \hat{w}_\nu} \, \bar{r}_{\hat{z}_\mu} \bar{r}_{\hat{w}_\nu} \right)-M}{M\qty[2\sum_{\mu}\sum_{z,w \in \{x, p\}} \left( \sigma_{\hat{z}_\mu \hat{w}_\mu}^2 + 2\, \sigma_{\hat{z}_\mu \hat{w}_\mu} \, \bar{r}_{\hat{z}_\mu} \bar{r}_{\hat{w}_\mu} \right) - M]}. 
\end{align}
which proves Theorem~\ref{Thm:privacy-n}.

\section{No complete average-phase privacy for Gaussian probes}
\label{App:no-rank-one-average-gaussian}

In this appendix we prove Theorem~\ref{thm:no-rank-one-average-gaussian-main}. For convenience, we first restate the theorem in a form adapted to the proof:

\begin{thmrestated}[No rank-one average-phase QFIm for Gaussian probes]
\label{thm:no-rank-one-average-gaussian}
Let $\hat\rho$ be an $M$-mode Gaussian state with $M\ge 3$, finite first moments $\bar{\v r}\in\mathbb R^{2M}$ and covariance matrix $\v\sigma\in\mathbb R^{2M\times 2M}$. Consider the local phase-encoding family $\hat\rho(\v\theta) = \hat U(\v\theta)\hat\rho\,\hat U(\v\theta)^\dagger$ where $\hat U(\v\theta) = \prod_{\mu=1}^{M} e^{i\theta_\mu \hat n_\mu}$ with $\hat n_\mu=\hat a_\mu^\dagger\hat a_\mu$ and  $\v\theta=(\theta_1,\dots,\theta_M)$. Let $\v Q$ be the quantum Fisher information matrix of this family. If every direction orthogonal to the average-phase direction is unobservable, i.e.
\begin{equation}
\bigl\{\v u\in\mathbb R^M:\v 1^{\ms T}\v u=0\bigr\} \subseteq \ker \v Q, \quad \text{with} \quad \v 1=(1,\dots,1)^{\ms T},
\end{equation}
then $\v Q=\v 0$. Consequently, no finite-energy Gaussian probe can satisfy $\operatorname{rank}\v Q=1$ and $\operatorname{range}(\v Q)=\operatorname{span}\{\v 1\}$.  Equivalently, for $M\ge 3$ there is no Gaussian probe whose only nonzero local-phase sensitivity is the average phase $\v v_{\mathrm{avg}}^{\ms T}\v\theta$, with $\v v_{\mathrm{avg}}=\tfrac{\v 1}{\sqrt M}$. 
\end{thmrestated}

\begin{proof} First note that no generality is lost by evaluating the QFIm at $\v{\theta}=\v 0$. Since the local number operators commute, $\hat U(\v{\theta}_0+\v{\delta}) = \hat U(\v{\theta}_0)\hat U(\v{\delta})$, and hence $\hat\rho(\v{\theta}_0+\v{\delta}) = \hat U(\v{\theta}_0)\hat\rho(\v{\delta}) \hat U(\v{\theta}_0)^\dagger$. The QFIm is invariant under parameter-independent unitary conjugation, so $\v Q(\v{\theta}_0)=\v Q(\v 0)$. We may therefore work at $\v{\theta}=\v 0$. Next, assume that the subspace $\mathcal H_0:=\{\v u\in\mathbb R^M:\v 1^{\rm T}\v u=0\}$ is contained in $\ker\v Q$. This subspace consists of all directions orthogonal to the average-phase direction. We now prove the result in three steps.

\begin{enumerate}
\item \emph{Step 1: vanishing QFI in all relative directions gives a relative-phase symmetry.}
For any $\v u\in\mathbb R^M$, define
\begin{equation}
\hat G_{\v u}:=\sum_{\mu=1}^M u_\mu \hat n_\mu .
\end{equation}
Equivalently, consider the one-parameter curve through parameter space $\hat\rho_{\v u}(t):=\hat\rho(t\v u)$. This is not a restriction on the unknown parameters, but simply the directional
submodel used to test sensitivity along $\v u$. Since the number operators commute, $\hat\rho_{\v u}(t)=e^{it\hat G_{\v u}}\hat\rho e^{-it\hat G_{\v u}}$. The QFI of this one-parameter submodel is the obtained from the multiparameter
QFIm, $F_Q[\hat\rho_{\v u}]=\v u^{\ms T}\v Q\v u $. Thus, for every $\v u\in\mathcal H_0\subseteq\ker\v Q$, the QFI in the direction $\v u$ vanishes. By Lemma~\ref{Lem:equivalence}, in the regular Gaussian phase-estimation setting considered here this is equivalent to $\partial_{\v u}\hat\rho(\v 0)=0$. For the present unitary encoding, $\partial_{\v u}\hat\rho(\v 0)=i[\hat G_{\v u},\hat\rho]$. Hence $[\hat G_{\v u},\hat\rho]=0$ for all $\v u\in\mathcal H_0$. Since this holds for every zero-sum vector, and since the number operators commute, it exponentiates to
\begin{equation}\label{Eq:app-relative-invariance}
\hat U(\v\phi)\hat\rho\hat U(\v\phi)^\dagger=\hat\rho \quad \text{for all} \quad  \v\phi\in\mathbb R^M \quad \text{with}\quad  \v 1^{\ms T}\v\phi=0 .
\end{equation}

\item \emph{Step 2: relative-phase invariance constrains the Gaussian moments.} Let $\hat{\v r}= (\hat x_1,\hat p_1,\dots,\hat x_M,\hat p_M)^{\ms T}$ be the quadrature vector. A local phase shift acts in phase space through the block-diagonal symplectic rotation $\hat U(\v\phi)^\dagger \hat{\v r}\,\hat U(\v\phi) = \v S(\v\phi)\hat{\v r} \v S(\v\phi) =\bigoplus_{\mu=1}^{M} R(\phi_\mu)$, where
\begin{equation}
    R(t) =
    \begin{pmatrix}
        \cos t & -\sin t\\
        \sin t & \cos t
    \end{pmatrix}.
\end{equation}
Therefore the Gaussian moments transform as $\bar{\v r}\mapsto \v S(\v\phi)\bar{\v r}$ and $\v\sigma\mapsto \v S(\v\phi)\v\sigma \v S(\v\phi)^{\ms T}$. Using the invariance in Eq.~\eqref{Eq:app-relative-invariance}, we obtain
\begin{equation}
    \bar{\v r}
    =
    \v S(\v\phi)\bar{\v r} \quad \text{and}\quad
    \v\sigma
    =
    \v S(\v\phi)\v\sigma \v S(\v\phi)^{\ms T}
    \label{Eq:app-moment-invariance}
\end{equation}
for every zero-sum phase vector $\v\phi$. We write the first moments in mode blocks as $\bar{\v r} =(\bar{\v r}_1,\dots,\bar{\v r}_M)$ with $\bar{\v r}_\mu\in\mathbb R^2$, and decompose the covariance matrix into
$2\times 2$ blocks, 
\begin{equation}
    \v\sigma
    =
    \begin{pmatrix}
        \v\sigma_{11} & \cdots & \v\sigma_{1M}\\
        \vdots & \ddots & \vdots\\
        \v\sigma_{M1} & \cdots & \v\sigma_{MM}
    \end{pmatrix},
    \qquad
    \v\sigma_{\mu\nu}\in\mathbb R^{2\times 2}.
\end{equation}

First, we show that relative-phase invariance forces all first moments to vanish. For any mode $\mu$ and choose another mode $\ell\neq\mu$ and set $\phi_\mu=t, \phi_\ell=-t$ and $\phi_\kappa=0$ otherwise. This gives a zero-sum phase vector, so Eq.~\eqref{Eq:app-moment-invariance} implies $\bar{\v r}_\mu = R(t)\bar{\v r}_\mu$ for all $t\in\mathbb R$. Differentiating at $t=0$ gives $\v J\bar{\v r}_\mu=0$. Because $\v J$ is invertible, $\bar{\v r}_\mu=\v 0$. As this holds for every mode $\mu$, we conclude that $\bar{\v r}=\v 0$. Second, every diagonal covariance block is isotropic. Using the same zero-sum phase vector as above, Eq.~\eqref{Eq:app-moment-invariance} implies $\v\sigma_{\mu\mu} = R(t)\v\sigma_{\mu\mu}R(t)^{\ms T}$ for all $t\in\mathbb R$. Differentiating at $t=0$ gives $\v J\v\sigma_{\mu\mu}-\v\sigma_{\mu\mu}\v J=0$. Writing the symmetric block as
\begin{equation}
  \v\sigma_{\mu\mu} = \begin{pmatrix} a & b\\ b & c \end{pmatrix},
\end{equation}
one finds
\begin{equation}
\v J\v\sigma_{\mu\mu}-\v\sigma_{\mu\mu}\v J = \begin{pmatrix} -2b & a-c\\ a-c & 2b \end{pmatrix}.
\end{equation}
Thus $b=0$ and $a=c$, so $\v\sigma_{\mu\mu} = \nu_\mu \iden_2$ for some $\nu_\mu\ge \tfrac12$. Third, we show that relative-phase invariance forces all off-diagonal covariance blocks to vanish. For any distinct modes $\mu\neq\nu$, choose a third mode $\ell\neq\mu,\nu$ and set $\phi_\mu=t$, $\phi_\ell=-t$, and $\phi_\kappa=0$ otherwise.  This again gives a zero-sum phase vector, while mode $\nu$ is left unrotated. Equation~\eqref{Eq:app-moment-invariance} then implies
$\v\sigma_{\mu\nu}=R(t)\v\sigma_{\mu\nu}$ for all $t\in\mathbb R$. Differentiating at $t=0$ gives $\v J\v\sigma_{\mu\nu}=0$. Because $\v J$ is invertible $\v\sigma_{\mu\nu}=0$. Since this holds for every pair of distinct modes $\mu\neq\nu$, all off-diagonal covariance blocks vanish. Combining the diagonal and off-diagonal constraints, the covariance matrix must have the form
\begin{equation}
\v\sigma = \bigoplus_{\mu=1}^{M}\nu_\mu \iden_2 \quad \text{and} \quad \bar{\v r}=\v 0 .
    \label{Eq:app-product-thermal-form}
\end{equation}

\item \emph{Step 3: the state is then invariant under all local phase shifts.} The form in Eq.~\eqref{Eq:app-product-thermal-form} is invariant under arbitrary local phase rotations, not only under zero-sum rotations. For every $\v\theta\in\mathbb R^M$, we have $\v S(\v\theta)\bar{\v r}=\bar{\v r}$ and $\v S(\v\theta)\v\sigma\v S(\v\theta)^{\ms T}=\v\sigma$. Since a Gaussian state is completely determined by its first moments and covariance matrix, it follows that $\hat U(\v\theta)\hat\rho\hat U(\v\theta)^\dagger=\hat\rho$ for all $\v\theta\in\mathbb R^M$. Thus the encoded family is constant, so every local phase direction has zero QFI. Hence $\v Q=\v 0$.
\end{enumerate}
This proves that if all relative-phase directions lie in $\ker\v Q$, then the average-phase direction also lies in $\ker\v Q$. Hence $\ker\v Q=\mathbb R^M$, and therefore $\v Q=\v 0$. In particular, the only possible QFIm with $\mathcal H_0\subseteq\ker\v Q$ is the zero matrix, so a nonzero rank-one QFIm supported on $\operatorname{span}\{\v 1\}$ cannot arise for an $M$-mode Gaussian probe with $M\ge3$.
\end{proof}

\end{document}